\documentclass[a4paper,12pt]{article}
\usepackage{graphicx}
\usepackage{subfig}
\title{Measurement of the Higgs boson decay branching ratio to charm quarks at the ILC}
\author{Y. Banda, T. Lastovicka, A. Nomerotski \\Particle Physics Department, University of Oxford \\Denys Wilkinson Building, Keble Road \\Oxford, OX1 3RH}
\date{}
\begin{document}
\maketitle
\begin{abstract}
We present an analysis for the measurement of the Higgs Boson branching ratio BR(H $\rightarrow c\bar{c})$ for a light Standard Model-like Higgs boson produced at 250 GeV centre of mass energy at the International Linear Collider (ILC). The tools and technique used for the analysis are described and relative statistical uncertainities are determined.
\end{abstract}

\section{Introduction} 

The measurement of the Higgs absolute branching ratios to all possible species
is an important part of the ILC program, giving a precision test of the Standard
Model prediction that the Higgs boson couples to each
particle in proportion to its mass~\cite{Higgs}. These measurements also discriminate between different `Beyond the SM' scenarios~\cite{mssm1,mssm2}.

This study considers the Higgs decay mode to charm quarks which results in two- and four-jet final states. 
The expected SM Higgs boson branching ratio to charm quarks is equal to 3.09\% for $m_H$ = 120 GeV and one of the main difficulties of the analysis 
is to separate this signal from the background of Higgs decays to b-quarks which has substantially larger Br of 67.92\%. 
The selected decay modes exercise the tagging of charm quarks which is particularly sensitive to the vertex detector performance~\cite{ZH}.

\section{Event Generation and Detector Simulation}

\begin{figure}[htbp]
\begin{center}
\includegraphics[scale=0.65]{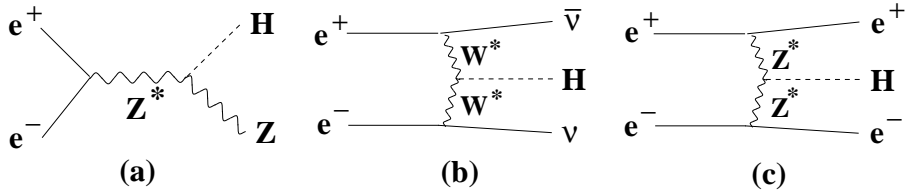}
\end{center}
\caption{Higgs production in (a)Higgs-strahlung, (b) WW-fusion and (c) ZZ-fusion processes}
\label{fig:higgsprod}
\end{figure}

\begin{figure}[htbp]
\begin{center}
\includegraphics[scale=0.40]{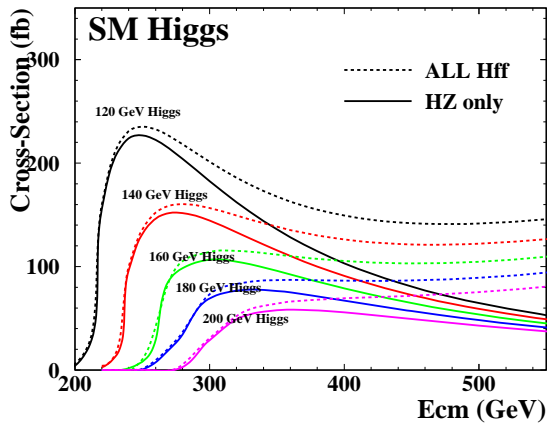}
\end{center}
\caption{Production cross-section for ZH process (solid lines) and all $Hf\bar{f}$ processes (dotted lines) for a 120-200 GeV SM Higgs as a function of 
the centre-of-mass energy}
\label{fig:zhxsec}
\end{figure}

The signal events comprise of Higgs bosons decaying into charm quarks produced in Higgs-strahlung, $e^{+}e^{-} {\to }$ ZH ignoring other contributing processes. The CP-even Higgs can also be produced via WW(ZZ)-fusion processes, 
$e^{+}e^{-} {\to } \nu\bar{\nu}W^{\ast}W^{\ast}{\to } \nu\bar{\nu}H (e^{+}e^{-} {\to } e^{+}e^{-}Z^{\ast}Z^{\ast} {\to } e^{+}e^{-}$H) as shown in Figure~\ref{fig:higgsprod}.
At low centre of mass energies the Higgs-strahlung process is dominant if $M_H \le 2M_W$ due to phase space considerations. The production cross-section is shown in Figure~\ref{fig:zhxsec}~\cite{acfa}. At larger CM energies ($\sqrt{s} \ge$ 500 GeV) the fusion processes start to dominate over the ZH process. The ZZ-fusion process is suppressed
by one order of magnitude compared to WW-fusion due to the ratio between neural and charged currents~\cite{lep2}. 

Both the signal and background events are produced at the centre of mass energy $\sqrt{s}=250$~GeV, total integrated luminosity of
250~fb$^{-1}$ and the Higgs boson is assumed to have mass 120 GeV. The choice of energy in this analysis maximizes the cross-section value for 
Higgs-strahlung. Standard Model events (mainly WW, ZZ and qq pairs) and Higgs decays to other fermions other than charm quarks are considered as backgrounds.  All 0, 2 and 4 fermion final states were generated using the Whizard Monte Carlo Event Generator~\cite{whiz}. 
PYTHIA~\cite{pyth} was used for the final state QED and QCD parton showering, fragmentation and decay to provide final-state 
observable particles. Photons from beamstrahlung and initial state radiation were also included in the simulations.

For this study, event samples are created conforming to the expected ILC baseline parameters of $\pm$80\% electron and $\mp$30\% positron polarization. 
About 7M events were processed through the full detector simulation, with individual events weighted to reflect the statistical sampling.

The detector response to generated events is simulated using the Geant4 toolkit~\cite{geant,geant2}, which provides the necessary classes
to describe the geometry of the detector, the transport and interaction of particles with materials and fields. A thin layer of 
Linear Collider specific code, SLIC~\cite{slic}, provides access to the Monte Carlo events, the detector geometry and 
the output of the detector hits. The detector parameters can be varied without having to rebuild the simulation executable binaries
since the geometries are fully described at runtime. The output is in the standard LCIO format~\cite{lcio} so that detectors using
other simulation packages can be analyzed and data generated using this system can be analyzed in other analysis frameworks. 

This analysis uses the Silicon Detector (sid02) model developed by the Silicon Detector (SiD) Design Study~\cite{sidloi}. The tracking, electromagnetic calorimetry, highly segmented hadronic calorimetry using PFA (Particle Flow Algorithm) and vertex detector are all silicon based. The SiD also incorporates a high field solenoid, iron flux return, and a muon identification system.
  
The vertex detector (VD) is especially important for this analysis since it relies on flavour tagging based on the lifetime information of decaying particles. The VD comprises of a central barrel system with 5 silicon pixel layers and forward and backward disk regions each with 4 pixel disks. It is enclosed within a double walled carbon fibre support. The pixels are 20 $\times \, 20 \mu m^2$ and the sensor thickness is taken to be 75$\mu$m. Sensor cut width is 8.68 mm in the innermost layer and 12.58 mm in all other layers. The sensor cut length is 125 mm for all layers. Carbon fibre cylinders support the sensors and the vertex detector is attached to the support tube with 25\% material coverage support disks. For SiD, the charged track momentum resolution is expected to be better than $\sigma(1/p_T) = 5\, \times 10^{-5} (GeV/c)^{-1}$ for high momentum tracks and the impact parameter resolution will be better than $\sigma_{r\phi} = \sigma_{rz} = 5\,  \oplus 10/(p \sin^{3/2}\theta$) [$\mu$m]. The overall material budget for the VD per layer is 0.15\% of the radiation length (X$_0$). Detailed geometry description of the detector is available at \underline{\textbf{http://lcsim.org/detectors/{\#}sid02}}.

\section{Analysis Tools}

Identification of jets is an importand part of this analysis. The fragmentation products of hadronic systems are forced in N jets.
The DURHAM algorithm~\cite{durham} is used in the analysis and provides y-cut values that determine the separation between 1 and 
2 jet cases and generally between N-1 and N jet cases. 

To provide the most probable kinematic configuration of the event topology a kinematic fitter (Marlin Kinfit~\cite{kinfit}) with four-momenta and mass
constraints is used. The fitter uses the method of Lagrange multipliers to determine the most probable value for the jet
four-momentum.

In order to identify primary, secondary and tertiary vertices the topological vertex finder ZVTOP is used. The algorithm 
is part of a vertexing package developed by the LCFI collaboration~\cite{lcfi}. It classifies events
on the number of found vertices and combines eight optimized variables for each type of event in a neural network, which is then
separately trained on samples of b-, c- and light quarks. The best discriminating variables are the corrected vertex mass, secondary 
vertex probability, impact parameter significance of the most significant track and the number of vertices in the event. There are
nine networks used with eight inputs, one hidden layer with 14 neurons and one output neuron.

\begin{figure}[htbp]
\begin{center}
\includegraphics[scale=0.45]{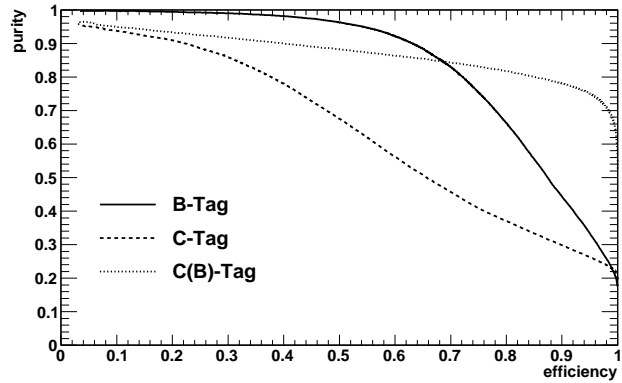}
\end{center}
\caption{Purity versus Efficiency for a 500 GeV di-jet sample for b-tagging c-tagging and c-tagging with b-only background.}
\label{fig:effpur}
\end{figure}

Figure~\ref{fig:effpur} shows the performance of the LCFI package optimized for the SiD detector. The plot shows the dependence of the purity on efficiency for the di-jet sample at $\sqrt{s}$ = 500 GeV for b-tagging, c-tagging and c-tagging with b-only background. The samples used for training of neural networks pass through the full SiD simulation and reconstruction.

The open source Fast Artificial Neural Network (FANN) package~\cite{fann} is used to provide neural networks for the final 
event selection. The package is written in C programming language called from within a C++ wrapper.

\section{Event Selection}

The analysis signature is dependent on the Z boson decay products (charged leptons, hadrons or neutrinos). The channels studied in this
analysis are the neutrino mode (Z decaying to neutrinos) and the hadronic mode (Z decaying to hadrons).

\begin{figure}[htbp]
\begin{center}
\subfloat[]{\includegraphics[scale=0.30]{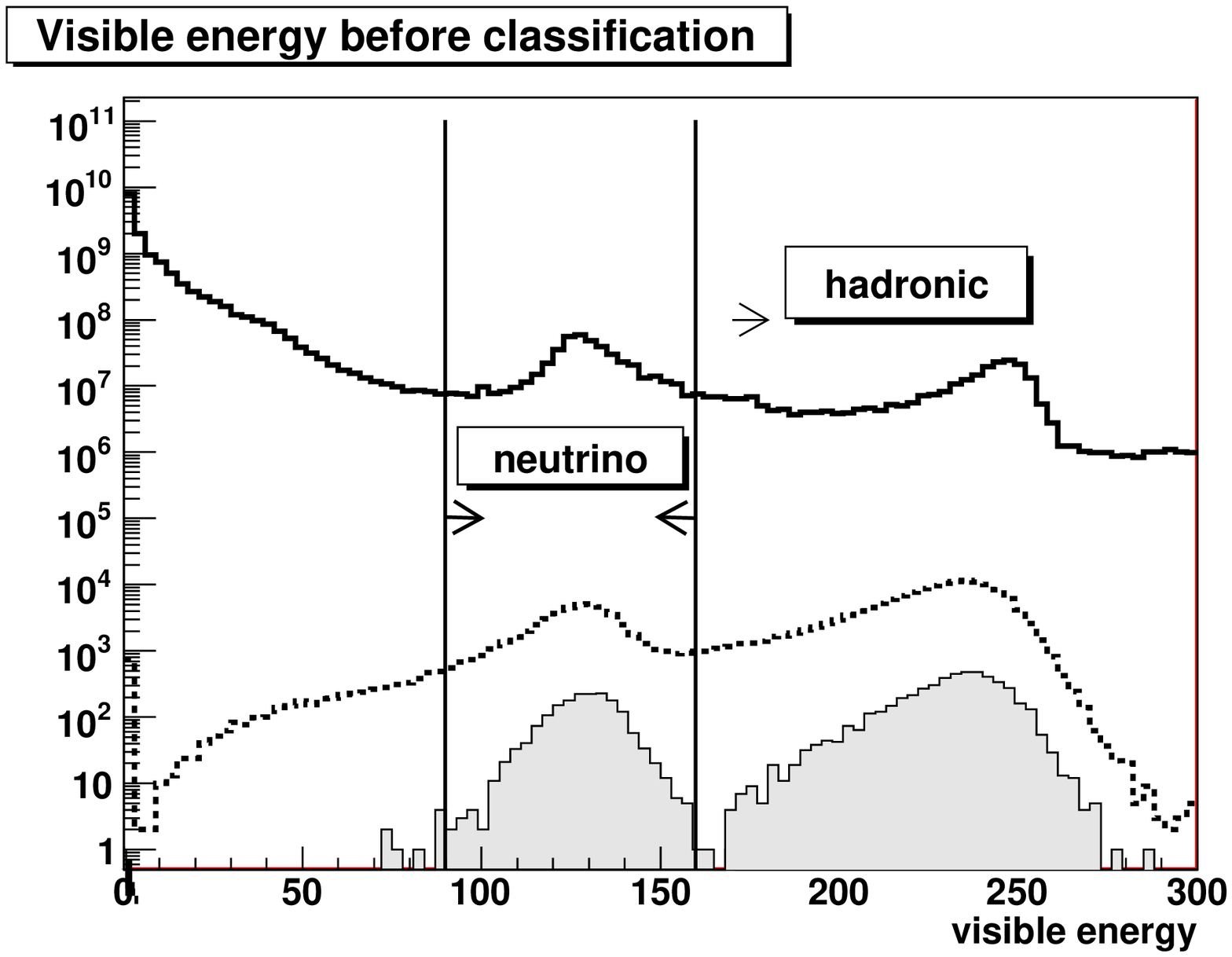}}
\subfloat[]{\includegraphics[scale=0.30]{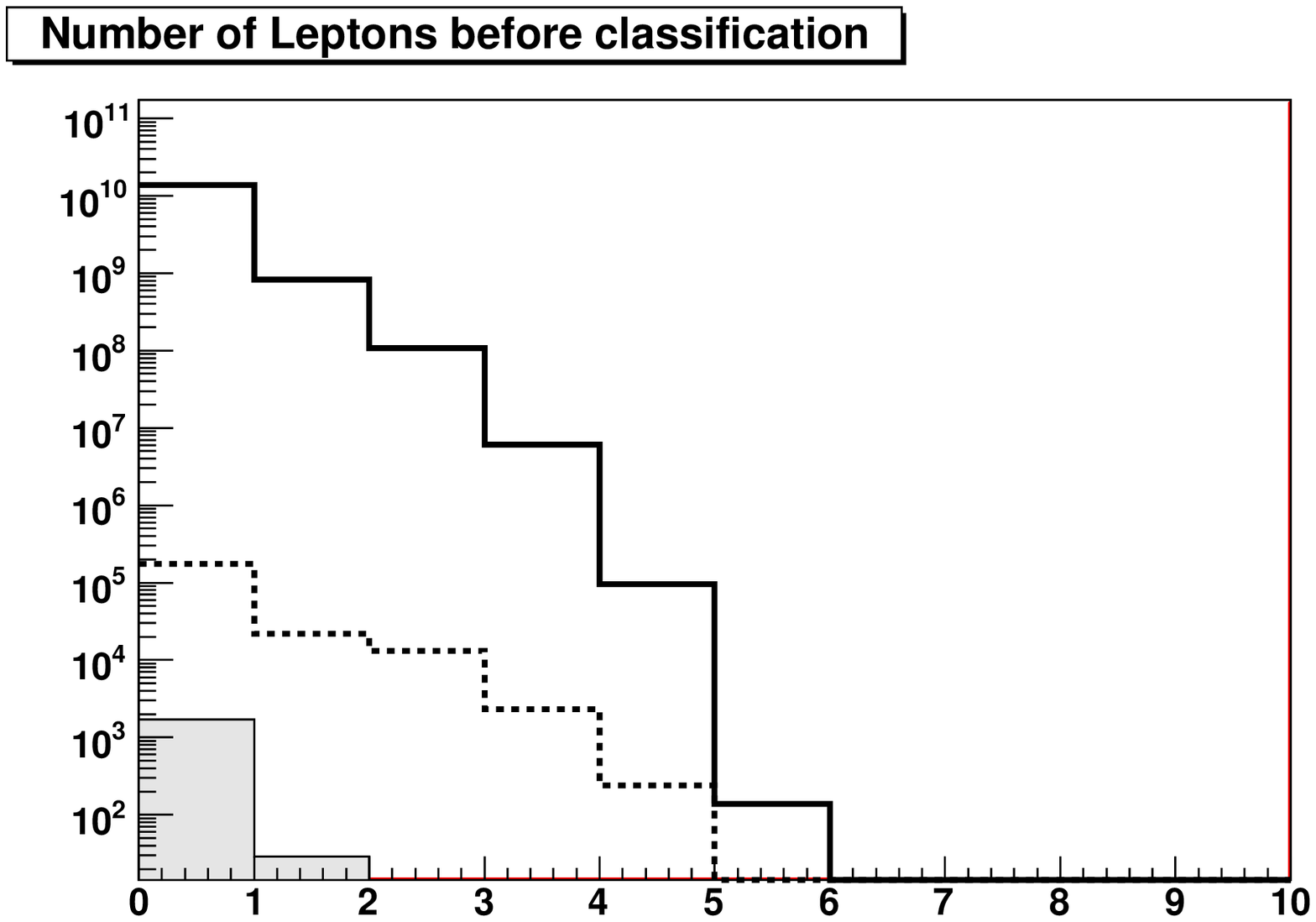}}
\end{center}
\caption{Visible energy (a) and the number of leptons per event (b) used for channel classification. Solid curves are SM background, dashed %%@
curves are inclusive Higgs sample and filled histograms are the signals.}
\label{fig:Enlep}
\end{figure}

The selection of signal events is performed in three stages. The first step involves the classification of events into two channels using the number of leptons and visible energy in the event. Visible energy is defined as the sum of energies of all reconstructed particles
in the event. Leptons are defined as reconstructed electrons or muons with momentum larger than $\:$15 GeV. Figure~\ref{fig:Enlep} shows the distributions of the visible energy and the number of leptons for the signal (filled histogram), Higgs background (dashed line) and SM background (solid line) before any selections. The neutrino mode is selected as events with no leptons and with visible energy in the $\:$ 90 to 160 GeV interval.
The hadronic mode is selected as events with no leptons and the visible energy above 170 GeV.

The second step is a cut based selection which reduces the backgrounds in the selected channel followed by a final stage using two neural
network (NN) discriminants trained on signal and two types of background samples, SM and Higgs.
The remaining events after the NN selection are used for the calculation of the cross sections and branching ratios.

\subsection{ The neutrino channel}

In this channel all reconstructed particles are clustered into two jets which, for
the signal, are assumed to come from the Higgs boson recoiling against two neutrinos from
the Z boson decay. The discrimination between the signal and background with a different number of jets is achieved by selection of the
$y_{min}$ parameter. It corresponds to the minimum y-parameter for the two jet hypothesis in the Durham algorithm. The y-parameter is defined as follows:
\begin{enumerate}
\item Given all pairs of particles $\textbf{i,j}$ calculate
\begin{equation}
y_{ij} = \frac{2min(E^{2}_i,E^{2}_j)(1-\cos\theta_{ij})}{E^{2}_{vis}},
\end{equation}
where $E_{vis}$ is the sum of the visible energies of all particles before any recombination.
\item If all $y_{ij}$ $>$ y-parameter stop calculation. The number of jets in the event is then defined as equal to the number of particles left.
\item Otherwise recombine the pair with the smallest $y_{ij}$ into a single particle.
\item Finally, go back to step 1.
\end{enumerate}
Therefore, in this channel the $y_{min}$ parameter is defined as the minimum value of the y-parameter that leads to a 2-jet event. Another powerful discriminant, the reconstructed invariant mass of two hadronic jets, is expected to be consistent with the Higgs mass. 

\begin{figure}[htbp]
\begin{center}
\subfloat[]{\includegraphics[scale=0.25]{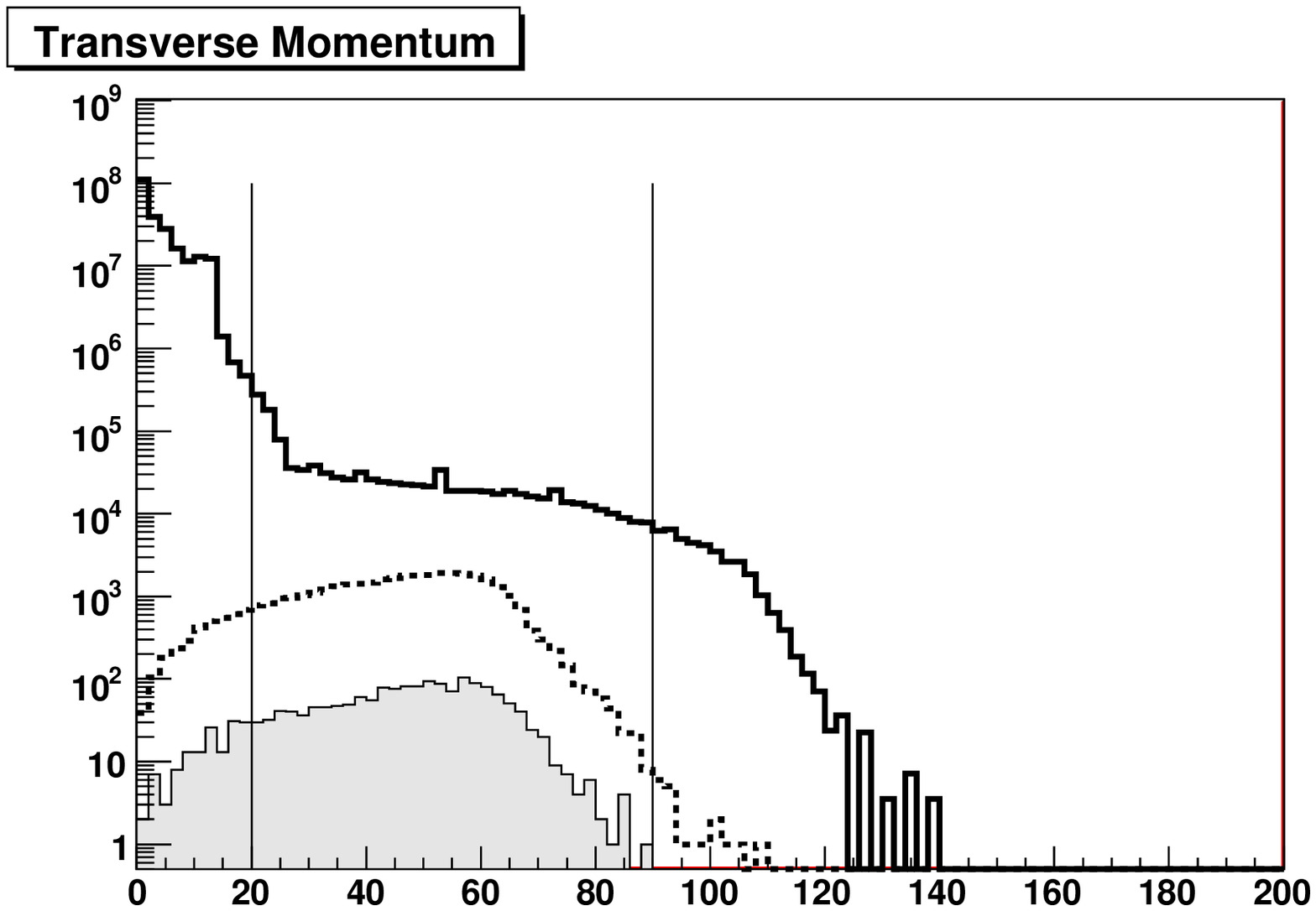}}
\subfloat[]{\includegraphics[scale=0.25]{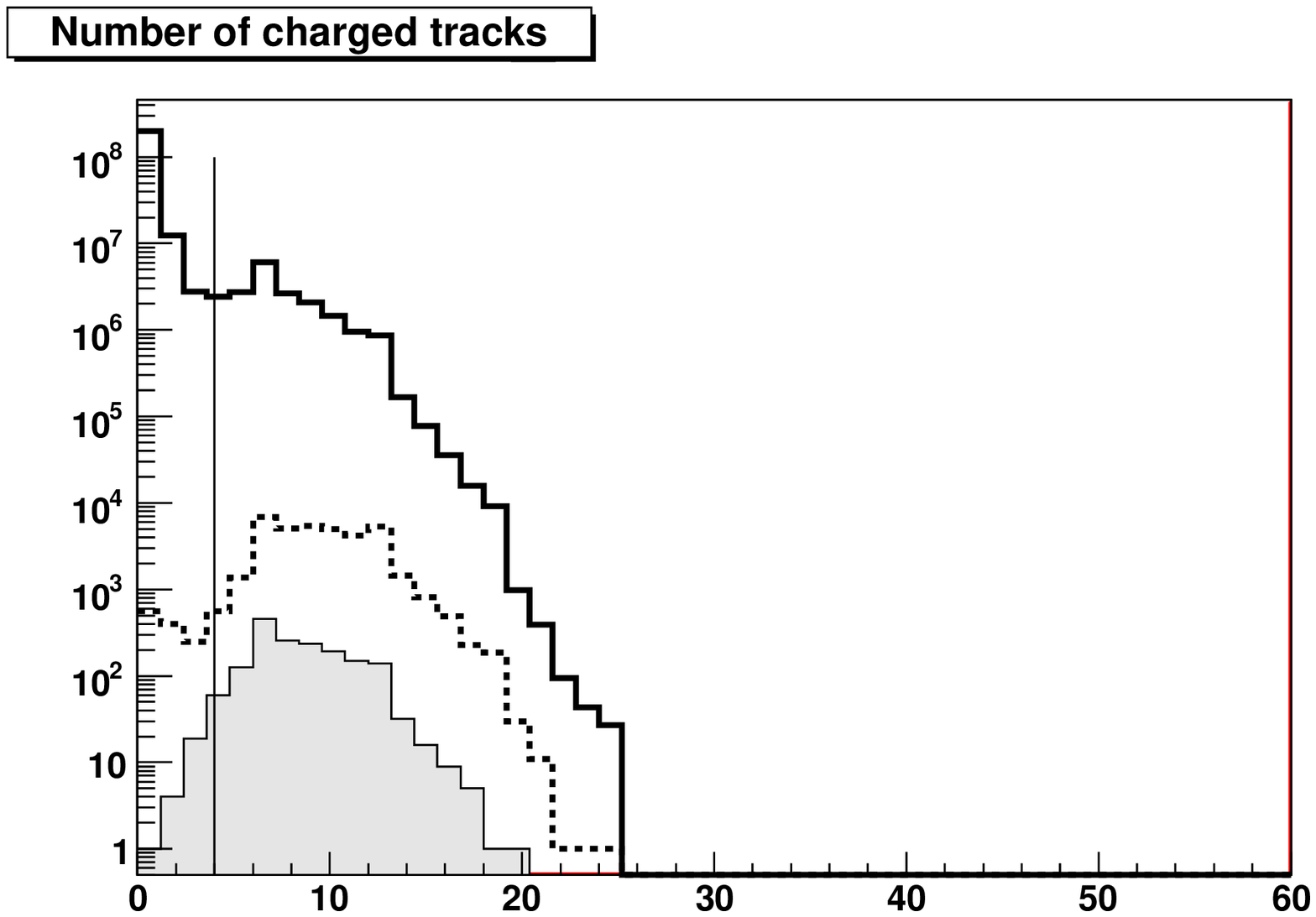}} \\
\subfloat[]{\includegraphics[scale=0.25]{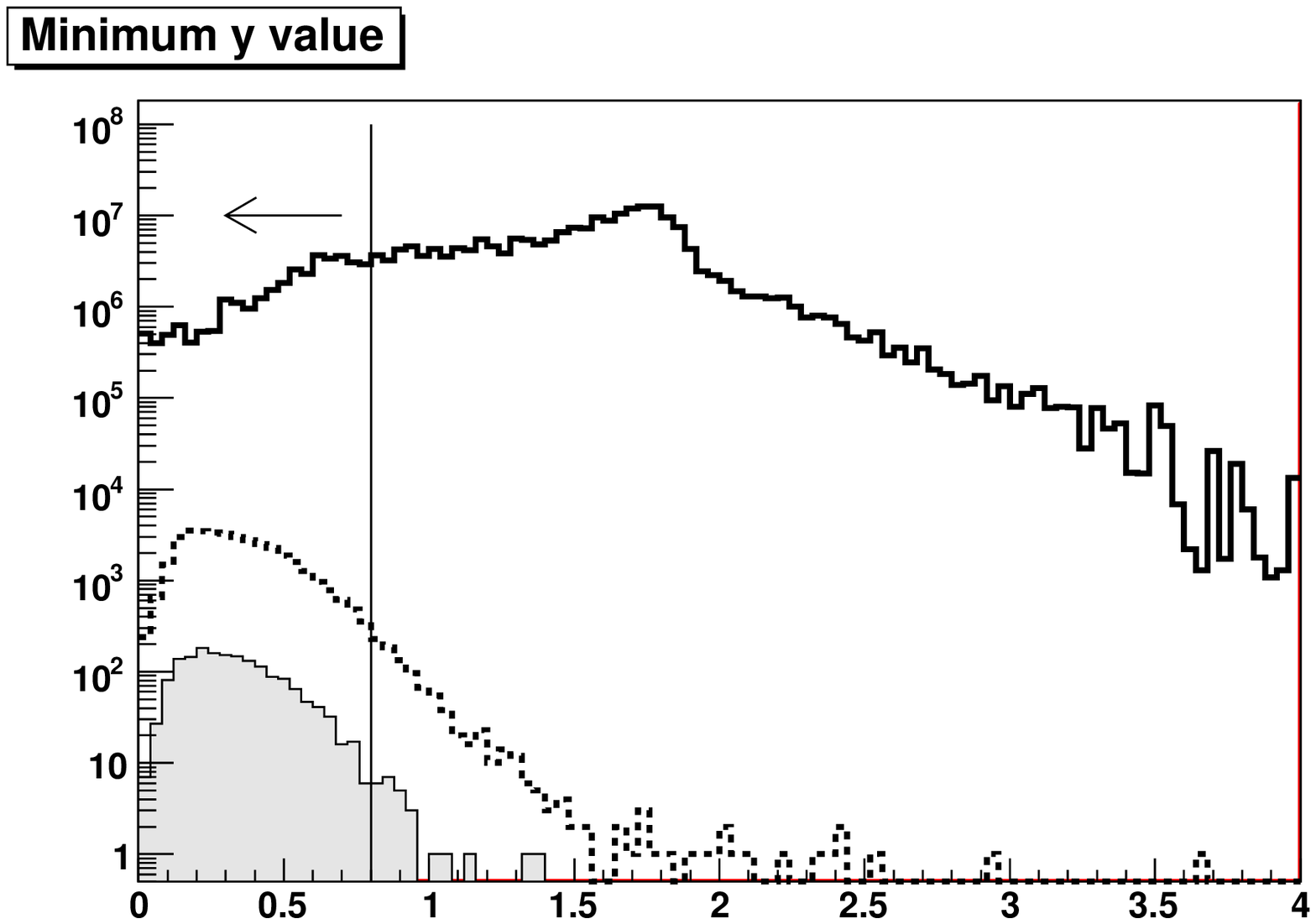}}
\subfloat[]{\includegraphics[scale=0.25]{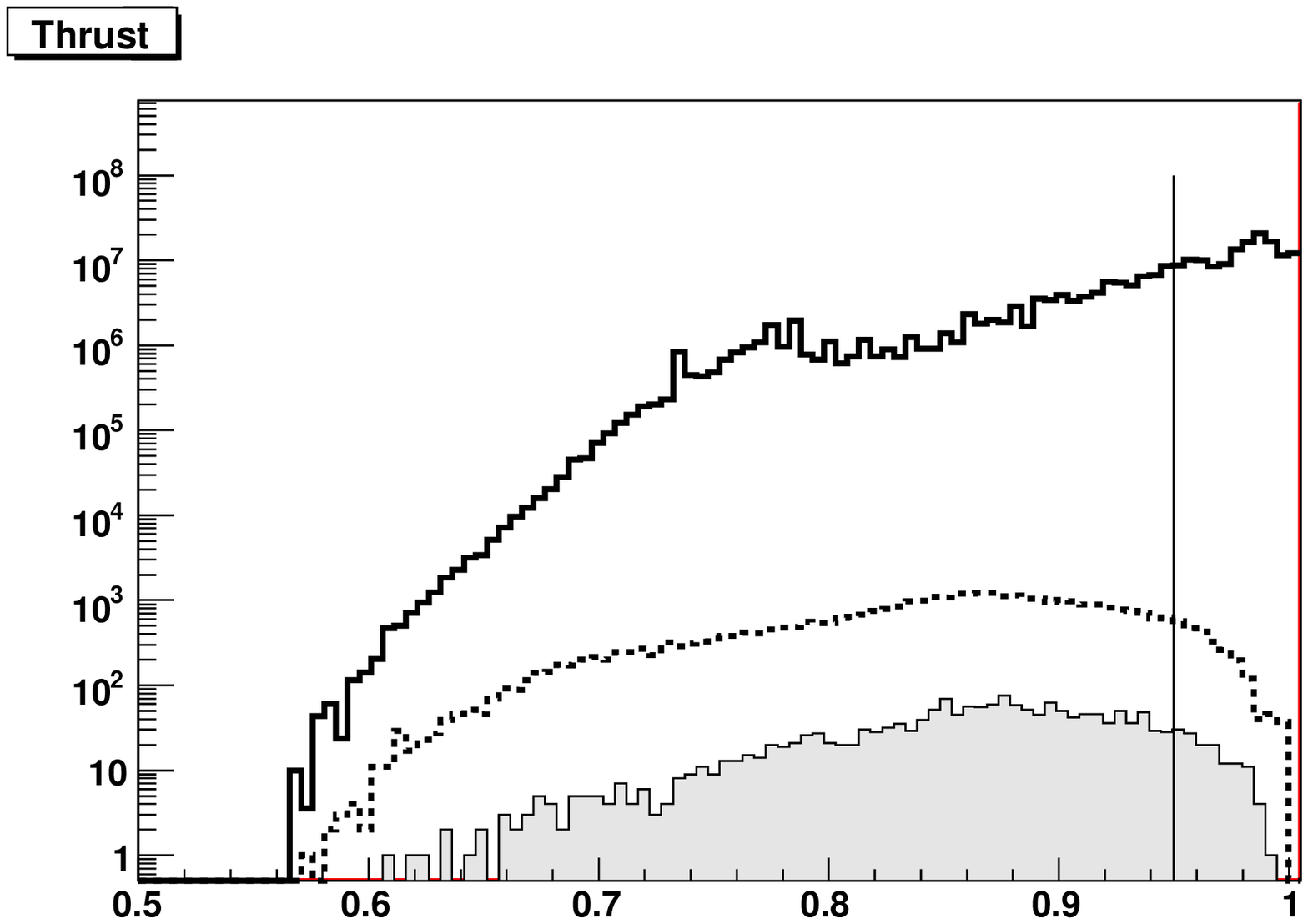}} \\
\subfloat[]{\includegraphics[scale=0.25]{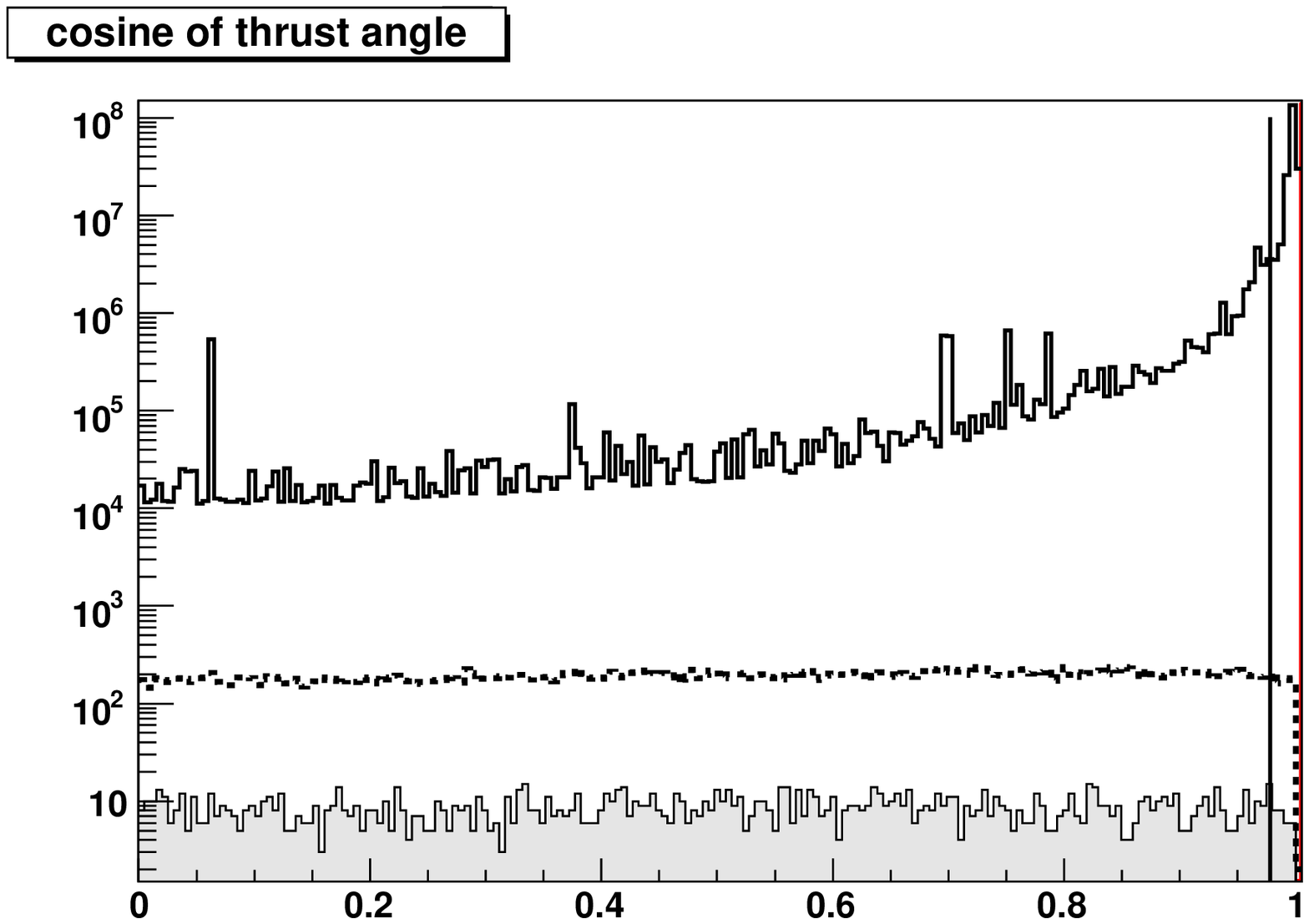}}
\subfloat[]{\includegraphics[scale=0.25]{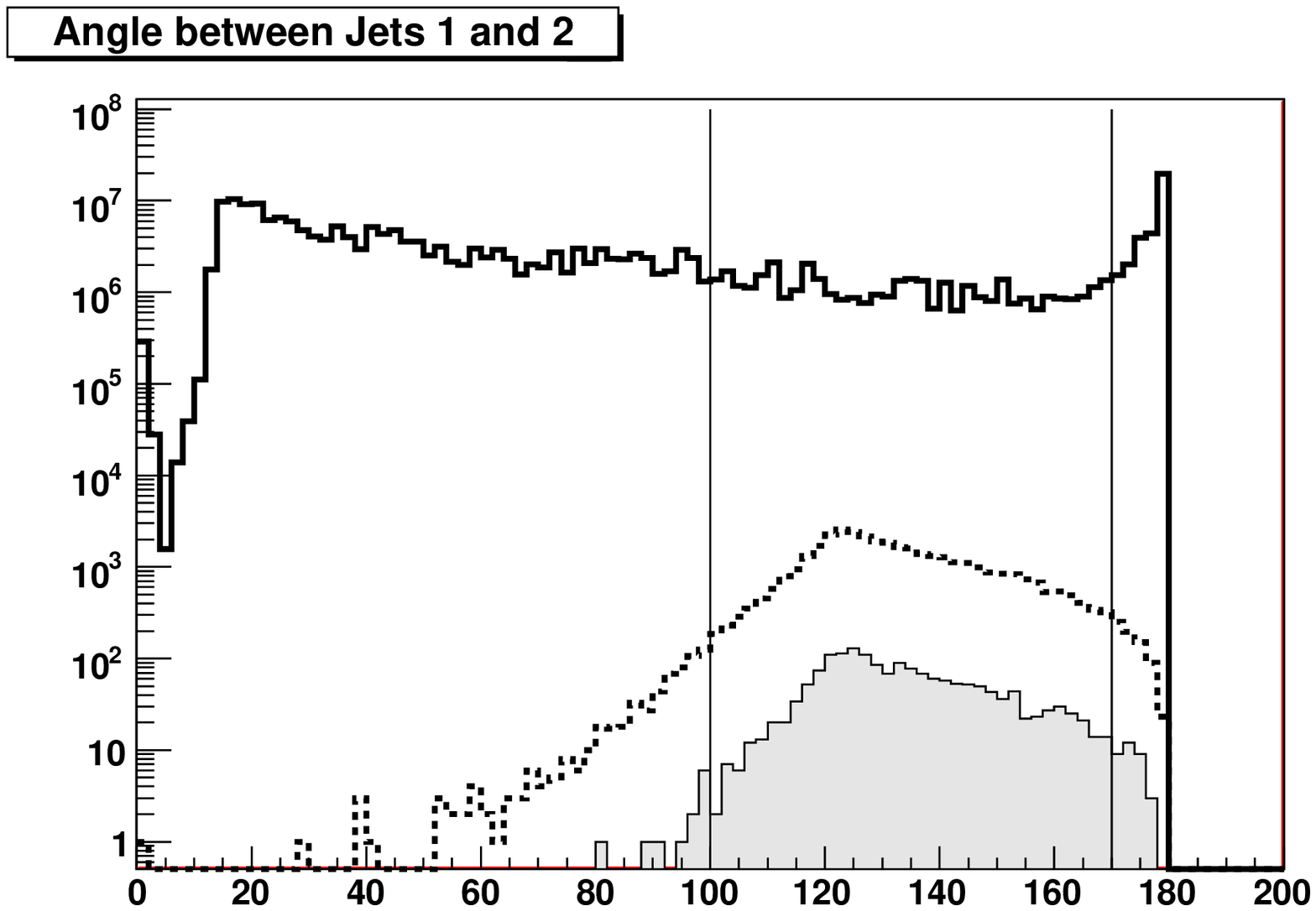}} \\
\subfloat[]{\includegraphics[scale=0.25]{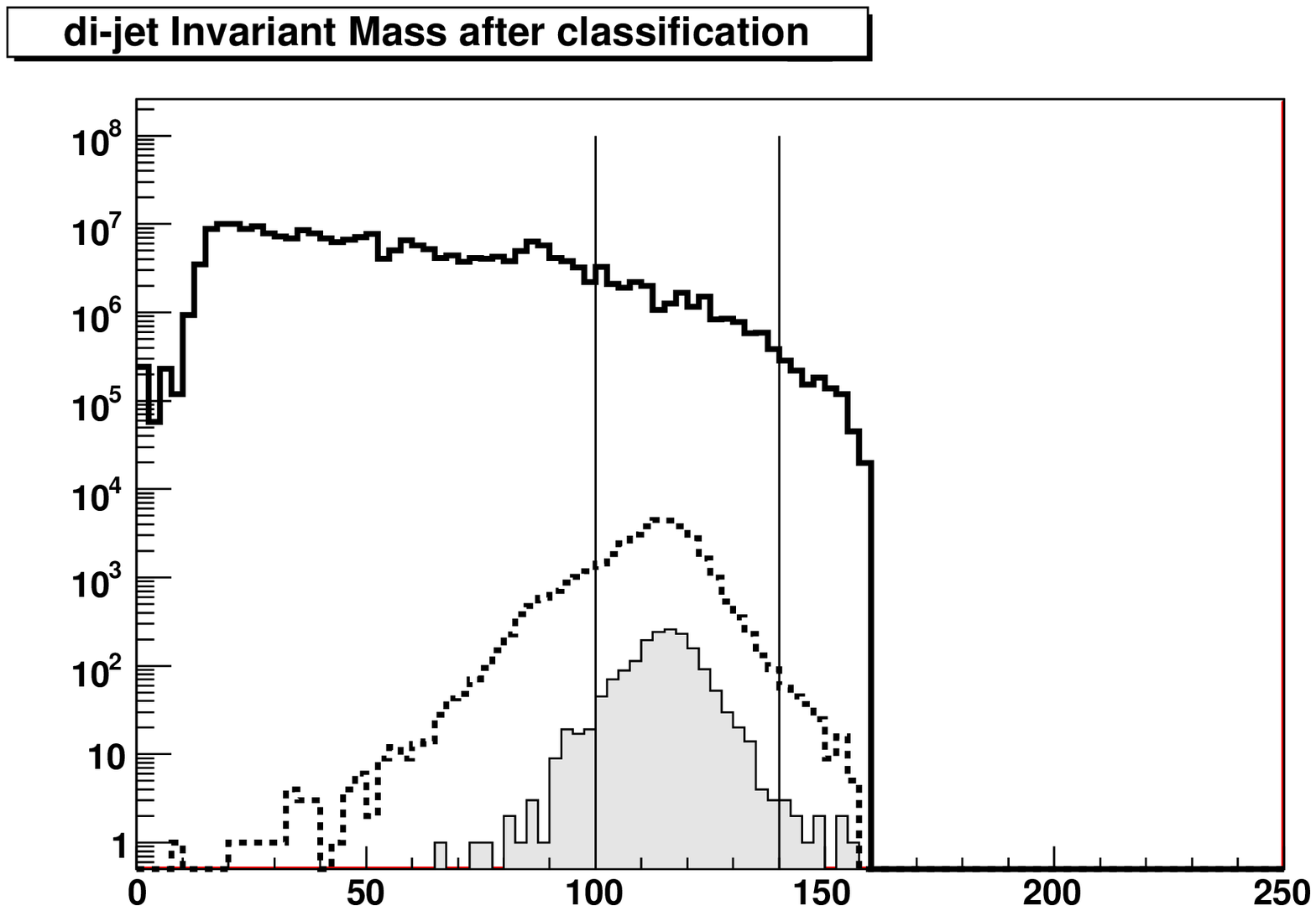}}
\subfloat[]{\includegraphics[scale=0.25]{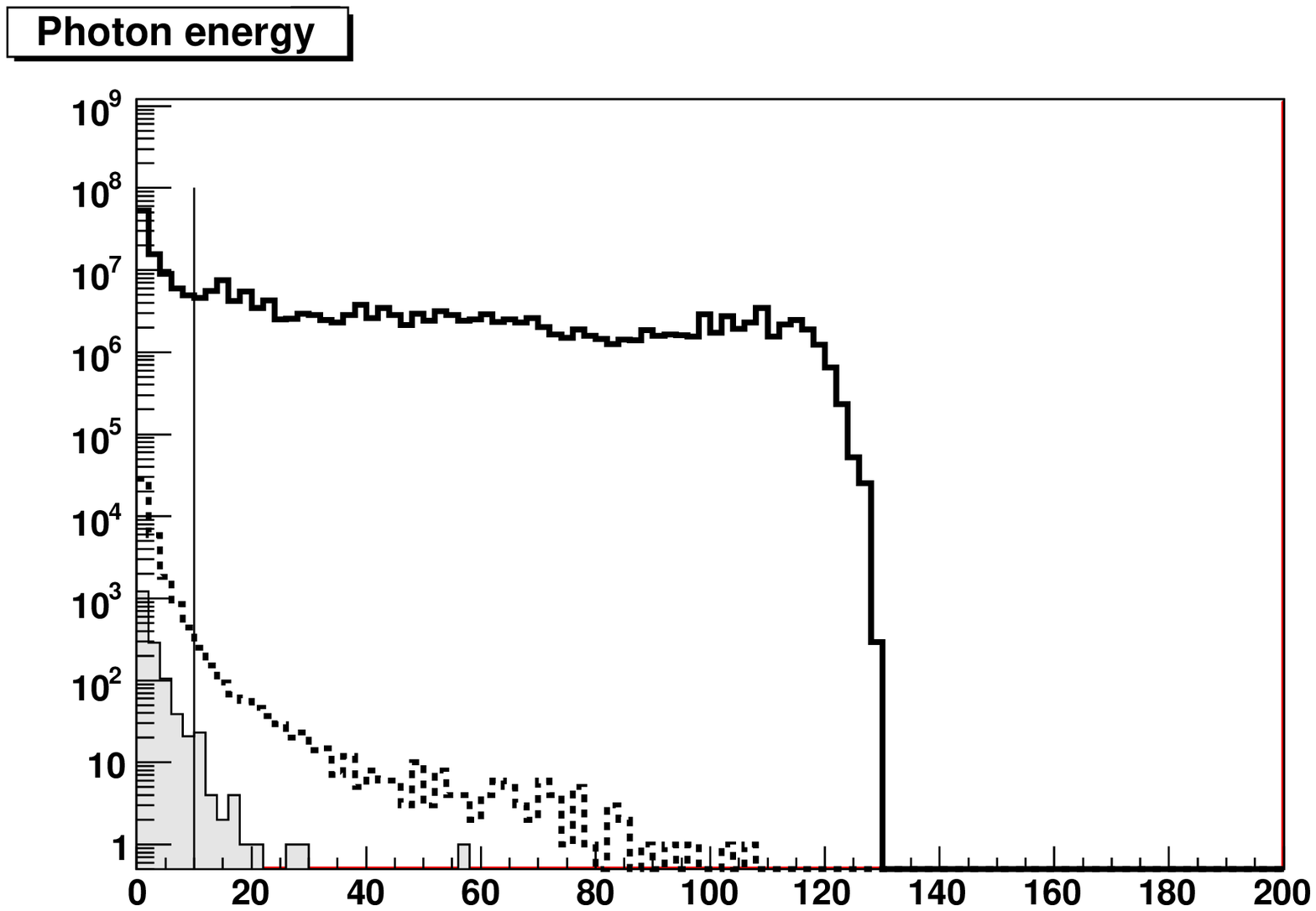}}
\end{center}
\caption{Distribution of pre-selection kinematic variables in the neutrino channel. Solid curves are SM background, dashed curves are inclusive Higgs %%@
sample and filled histograms is the signals.}
\label{fig:nu}
\end{figure}

Figure~\ref{fig:nu} shows distributions of various variables used for pre-selection for the signal and backgrounds after classification. The kinematic variables used are:
\begin{itemize}
\item Transverse momentum of jet, $P_T$: The majority of SM background events are soft compared to signal events. The signal peak is expected to be at 
$m_H/2$.
\item $n_{tracks}$ per jet: Require more than 4 charged tracks to reject purely leptonic events
\item $-\log(y_{min}$): This variable is used to exclude fully hadronic WW (and ZZ) events which are 4-jet events mis-identified as 2-jet events. The 
$-\log(y_{min}$) value is larger for the backgrounds compared to the signal.
\item Thrust: This is the maximum directed momentum given by
\begin{equation}
T \equiv \max_{\vec{n}} \frac{\displaystyle\sum_{i}\mid \vec{p_i}\cdot \vec{n} \mid}{\displaystyle\sum_{i}\mid \vec{p_i}\mid}
\end{equation}
Signal events are less boosted and are more spherical than the background events.
\item $\mid$$\cos(\theta_{thrust})$$\mid$: Signal events are produced more centrally in the detector while majority of the background processes 
have a strongly forward peaking angular distribution.
\item Angle between jets: A large fraction of background events have back-to-back jets but there also backgrounds with low angular separation 
due to low boost. Signal events are confined are confined to a range of angular values due to constraints coming from Z and H production.
\item Di-jet invariant mass: The signal di-jet mass is required to be between 100 and 140 GeV to reject low mass hadronic systems from WW, ZZ and two-photon events
\item Highest reconstructed photon energy: The signal events rarely have hard photons and this helps reject both highly energetic initial state radiation (ISR) photons and  hard photons from di-jet events which occur for the background. No photon isolation is required.
\end{itemize} 

The full list of selections is given in Table~\ref{tab:HZ1}. \\ \\

\begin{table}[h]
\centering
\begin{tabular}{lllll}
\hline Cut & && selection & value  \\
\hline
1. &  20 & $<$ & p$_T$ of jet & $<$ 90 GeV \\
2. & && number of charged tracks per jet & $>$ 4\\
3. & && $-\log(y_{min}$) & $<$ 0.8 \\
4. & && thrust & $< 0.95$ \\
5. & && $ \cos(\theta_{thrust}) $&$< 0.98$ \\
6. & $100^{\circ}$ &$<$ & angle between jets &$< 170^{\circ}$ \\
7. & 100 GeV &$<$ & di-jet invariant mass &$<$ 140 GeV \\
8. & && Highest reconstructed photon energy & $<$ 10 GeV \\
\hline
\end{tabular}
\caption{Selections for the neutrino channel}
\label{tab:HZ1}
\end{table}

The background includes all SM processes. The most important are 2-fermion events, ZZ pairs decaying to neutrinos and hadrons and WW pairs
where one W decays hadronically and the other W decays into a neutrino and a lepton which escapes undetected.
Table~\ref{tab:Nuevts} shows the number of events before and after pre-selection cuts.

\begin{table}[htpb]
\centering
\begin{tabular}{|c|c|c|c|}
\hline
Cuts & SM background & Higgs background & Signal \\ \hline
(i) Before Classification  & 9275594683 & 17475 & 652 \\ \hline
(0) After Classification & 45936973 & 14294 & 637 \\ \hline 
(1) & 18374789 & 13783 & 619 \\ \hline
(2) & 17123140 & 13729 & 618 \\ \hline
(3) & 6849256 & 13416 & 609\\ \hline
(4) & 685329 & 12179 & 560 \\ \hline
(5) & 627113 & 11945 & 550 \\ \hline
(6) & 576422 & 10226 & 525 \\ \hline
(7) & 203292 & 10088 & 519 \\ \hline
(8) & 109057 & 9902 & 506 \\ \hline
\end{tabular}
\caption{Number of $c\bar{c}$ events before channel classification, after channel classification and after pre-selection cuts in the neutrino mode.}
\label{tab:Nuevts}
\end{table}

\begin{figure}[htbp]
\begin{center}
\subfloat[]{\includegraphics[scale=0.30]{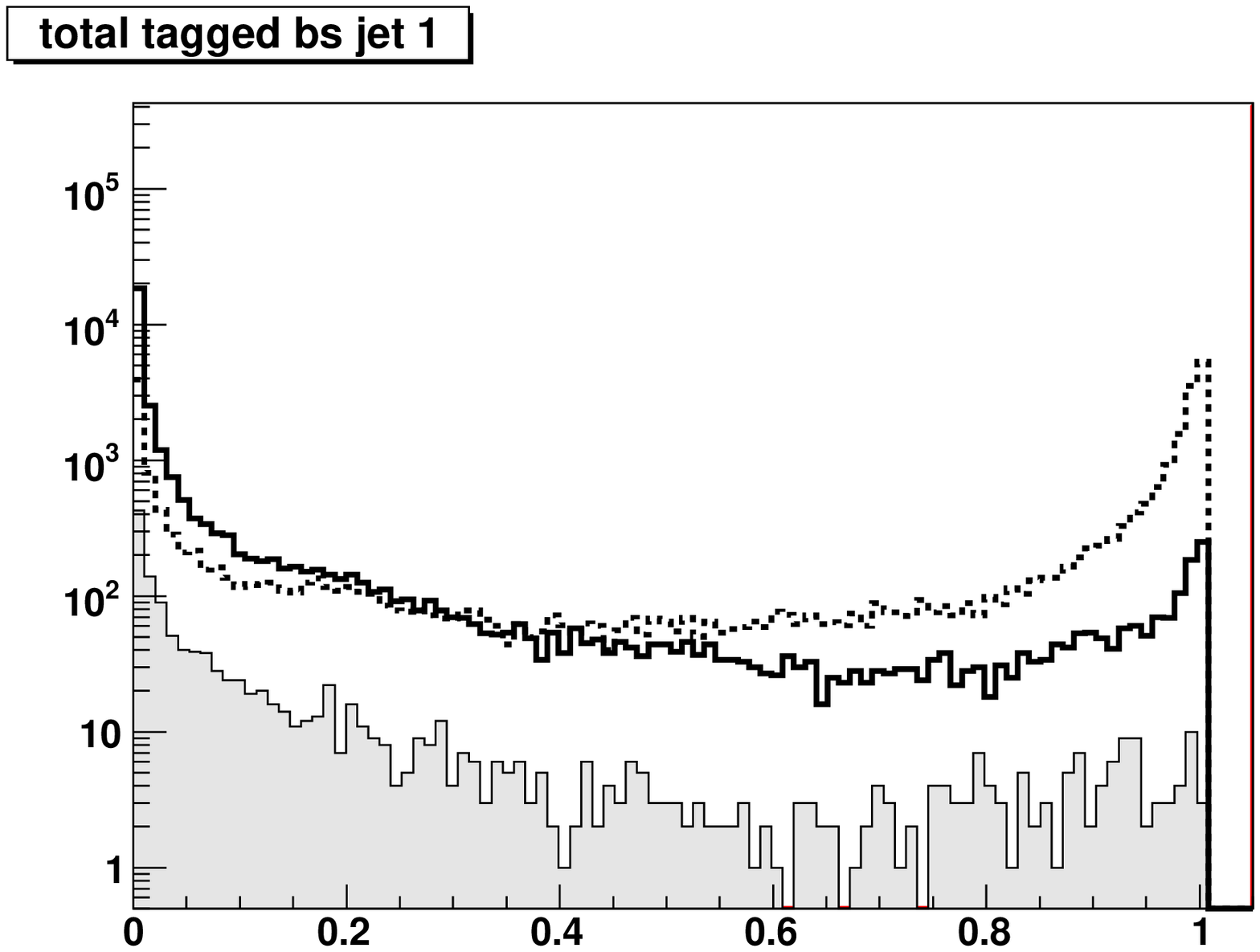}}
\subfloat[]{\includegraphics[scale=0.30]{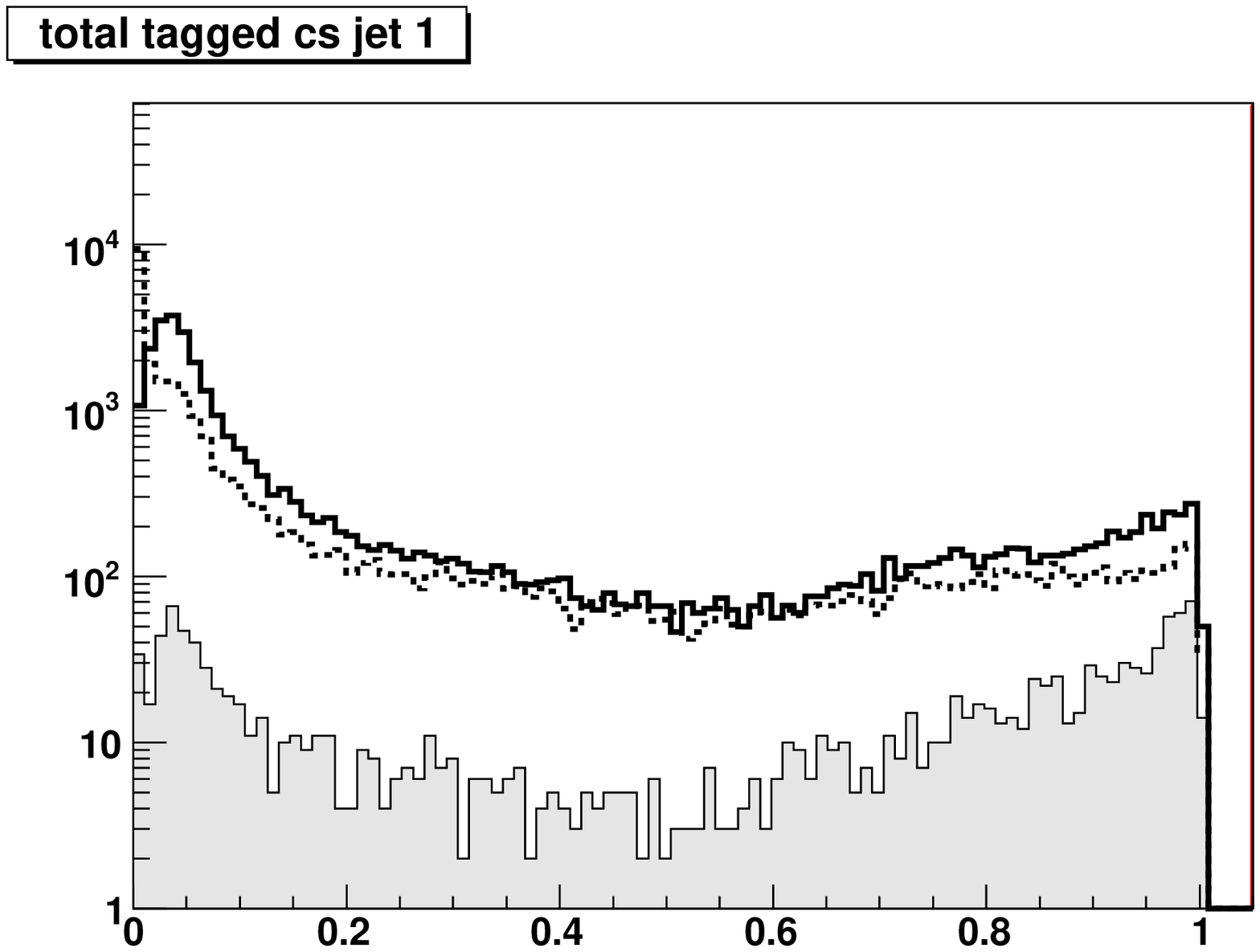}} \\
\subfloat[]{\includegraphics[scale=0.30]{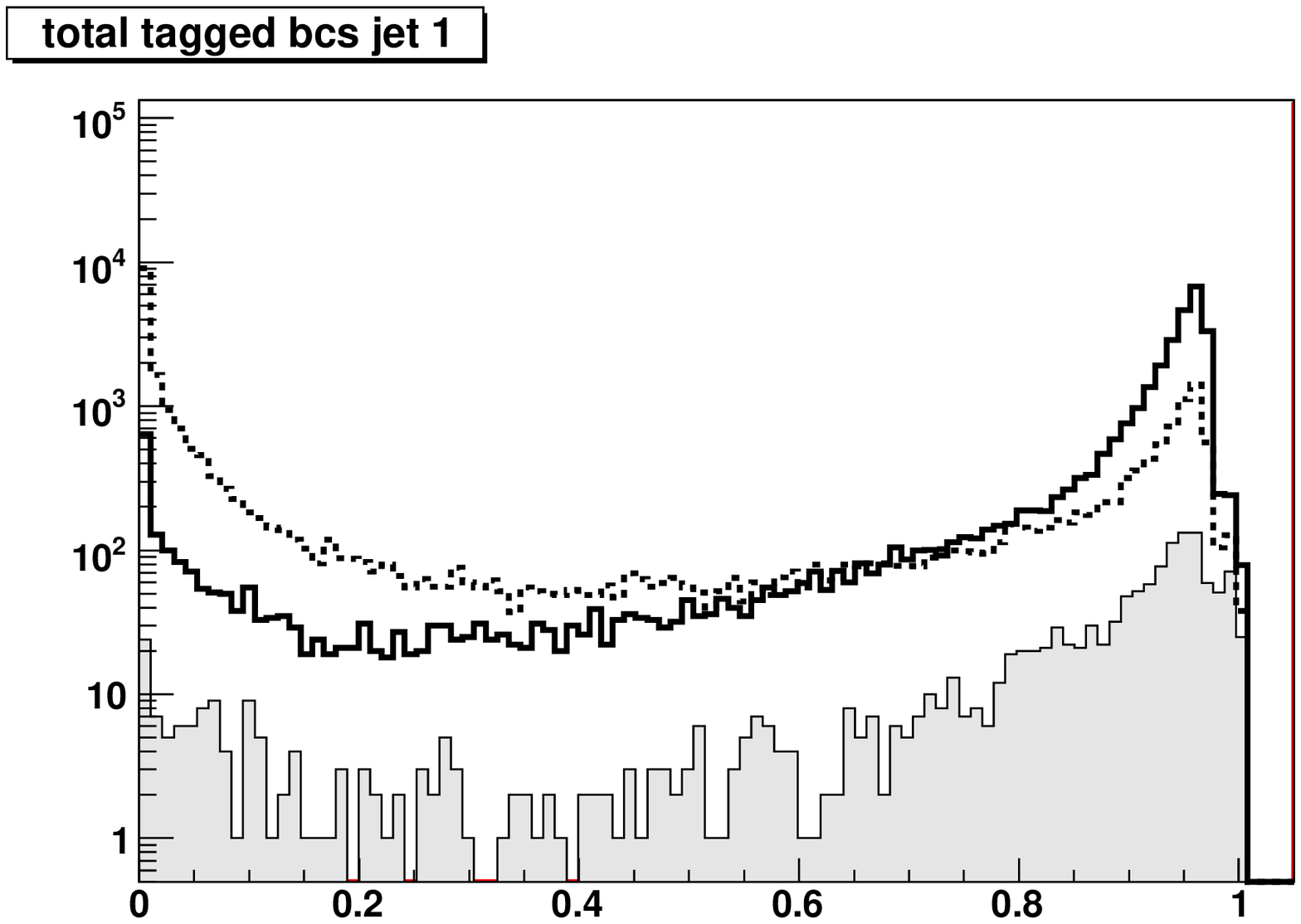}}
\end{center}
\caption{Neutrino channel: (a) b-tag; (b) c-tag; (c) c-tag with b background only;. Solid curves are SM 
background, dashed curves are Higgs background sample and filled histograms are the signal.}
\label{fig:ccNN}
\end{figure}

\begin{figure}[htbp]
\begin{center}
\subfloat[]{\includegraphics[scale=0.30]{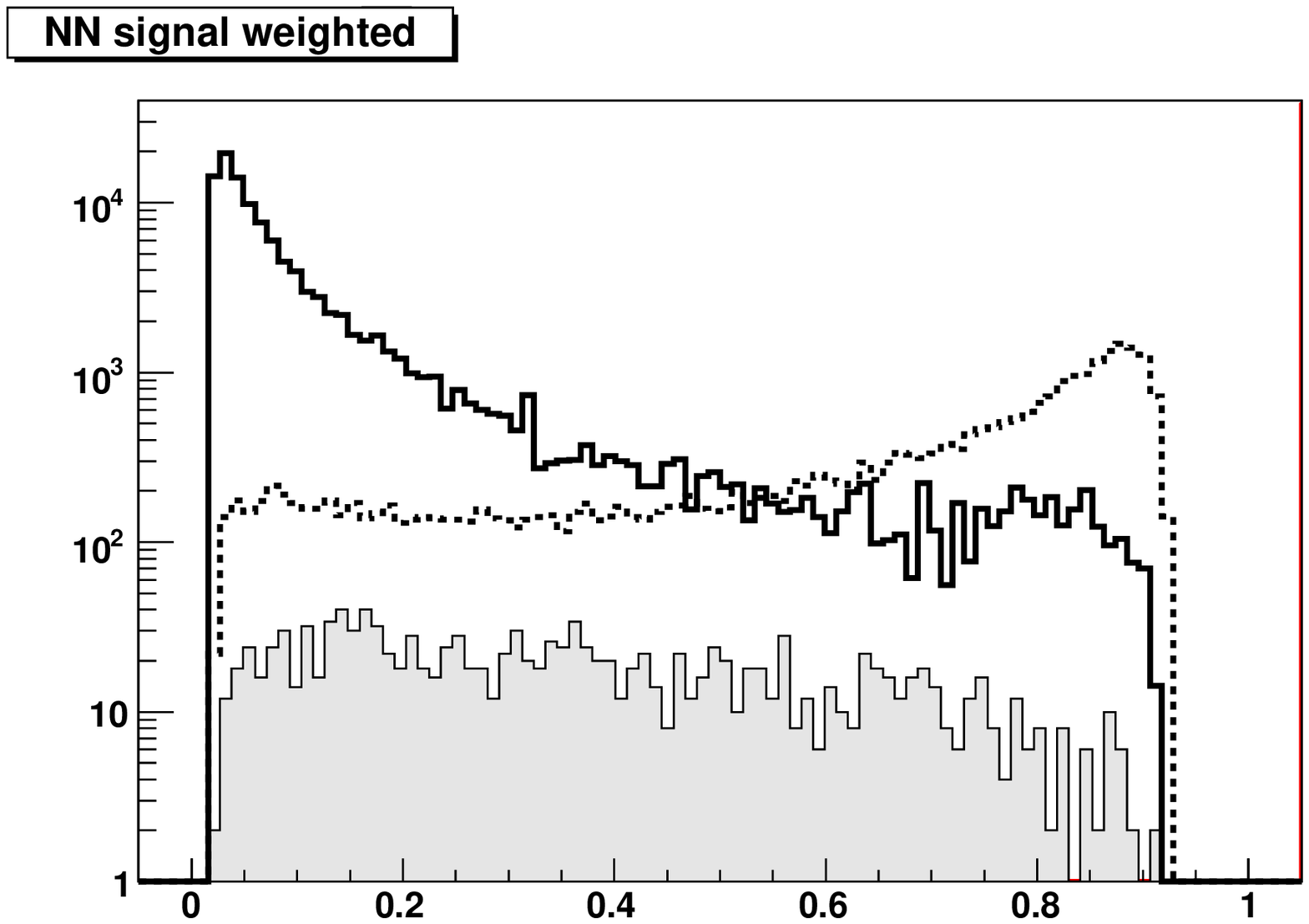}}
\subfloat[]{\includegraphics[scale=0.30]{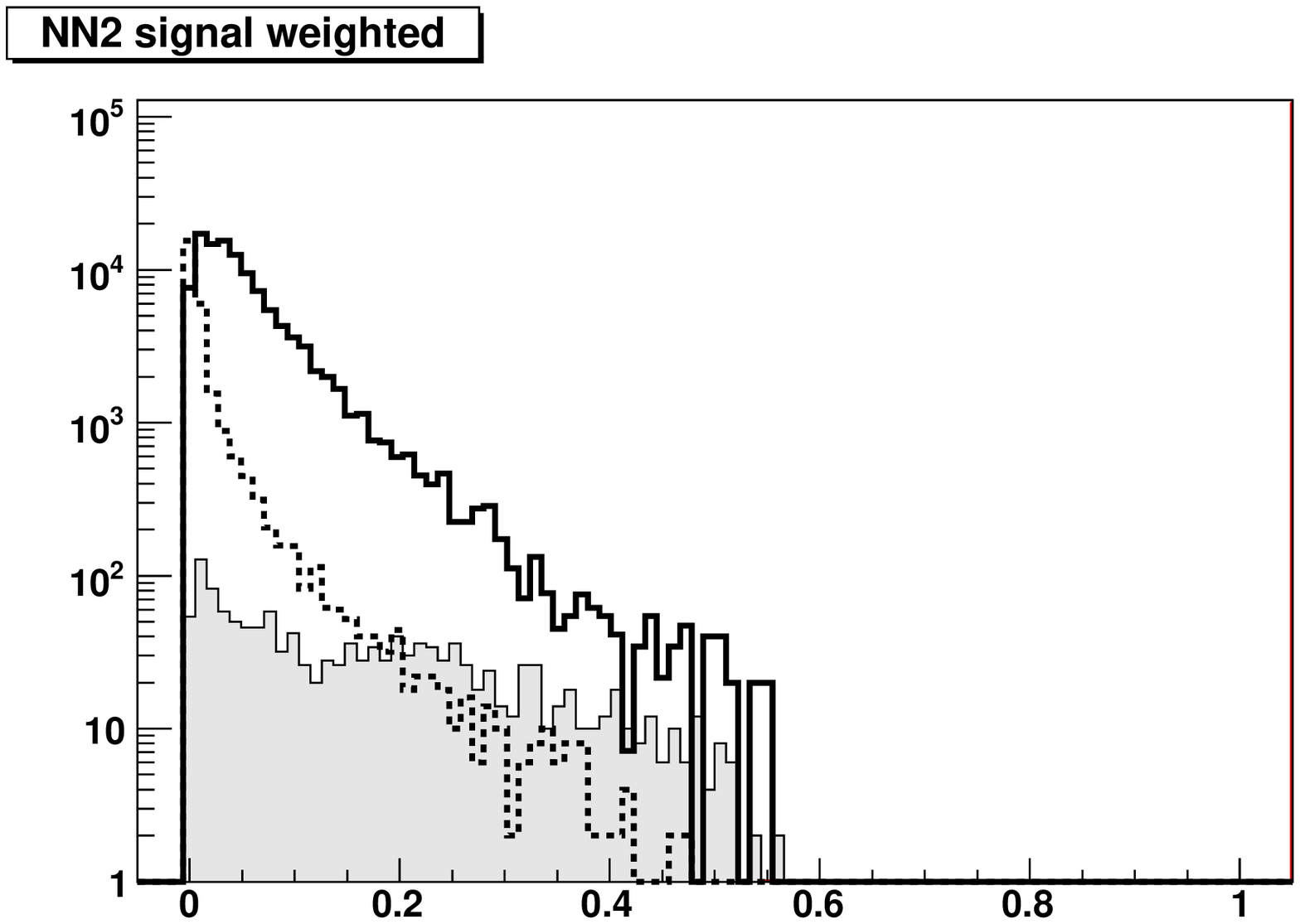}}
\end{center}
\caption{Neutrino channel: (a) NN$_{SM-Higgs}$ output; (b) NN$_{Higgs-signal}$ output. Solid curves are SM 
background, dashed curves are Higgs sample and filled histograms are the signals.}
\label{fig:nns}
\end{figure}

The remaining events are categorized using the neural networks implemented in FANN~\cite{fann}. The NN input variables include all the variables stated
above with addition of the LCFI flavour tag outputs for both jets. Figure~\ref{fig:ccNN} shows the distribution of the three possible
LCFI flavour tags, `b-tag', `c-tag' and  `c-tag with b background only' for the leading jet. The first NN is trained to
distinguish the SM background from the inclusive Higgs sample and to produce the NN$_{SM-Higgs}$ output; and, the second NN to distinguish
the signal from the Higgs background sample and to produce the NN$_{Higgs-signal}$ output. The outputs of the trained NNs are shown in Figure~\ref{fig:nns}.

\begin{figure}[htbp]
\begin{center}
\subfloat[]{\includegraphics[scale=0.30]{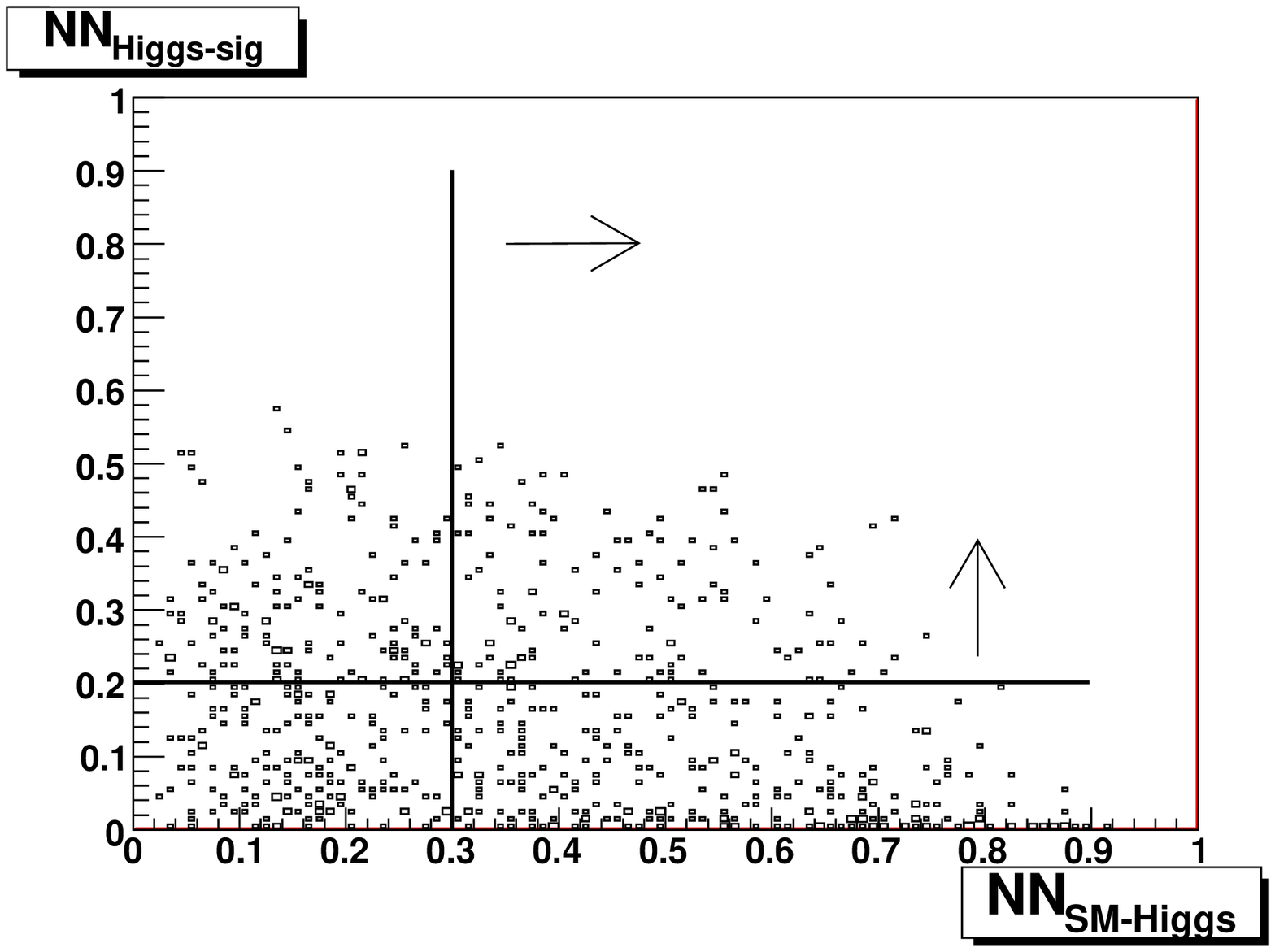}}
\subfloat[]{\includegraphics[scale=0.30]{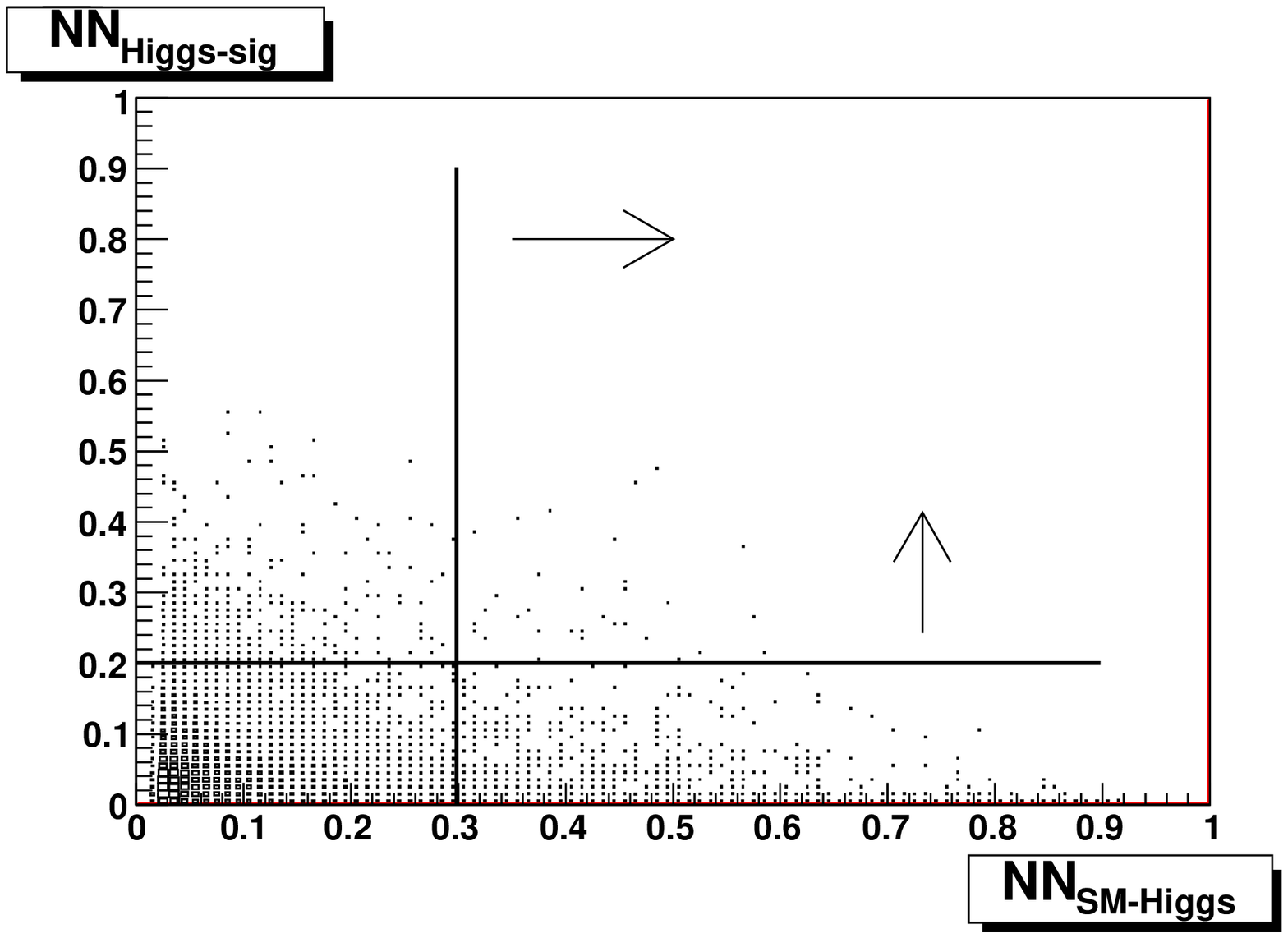}} \\
\subfloat[]{\includegraphics[scale=0.30]{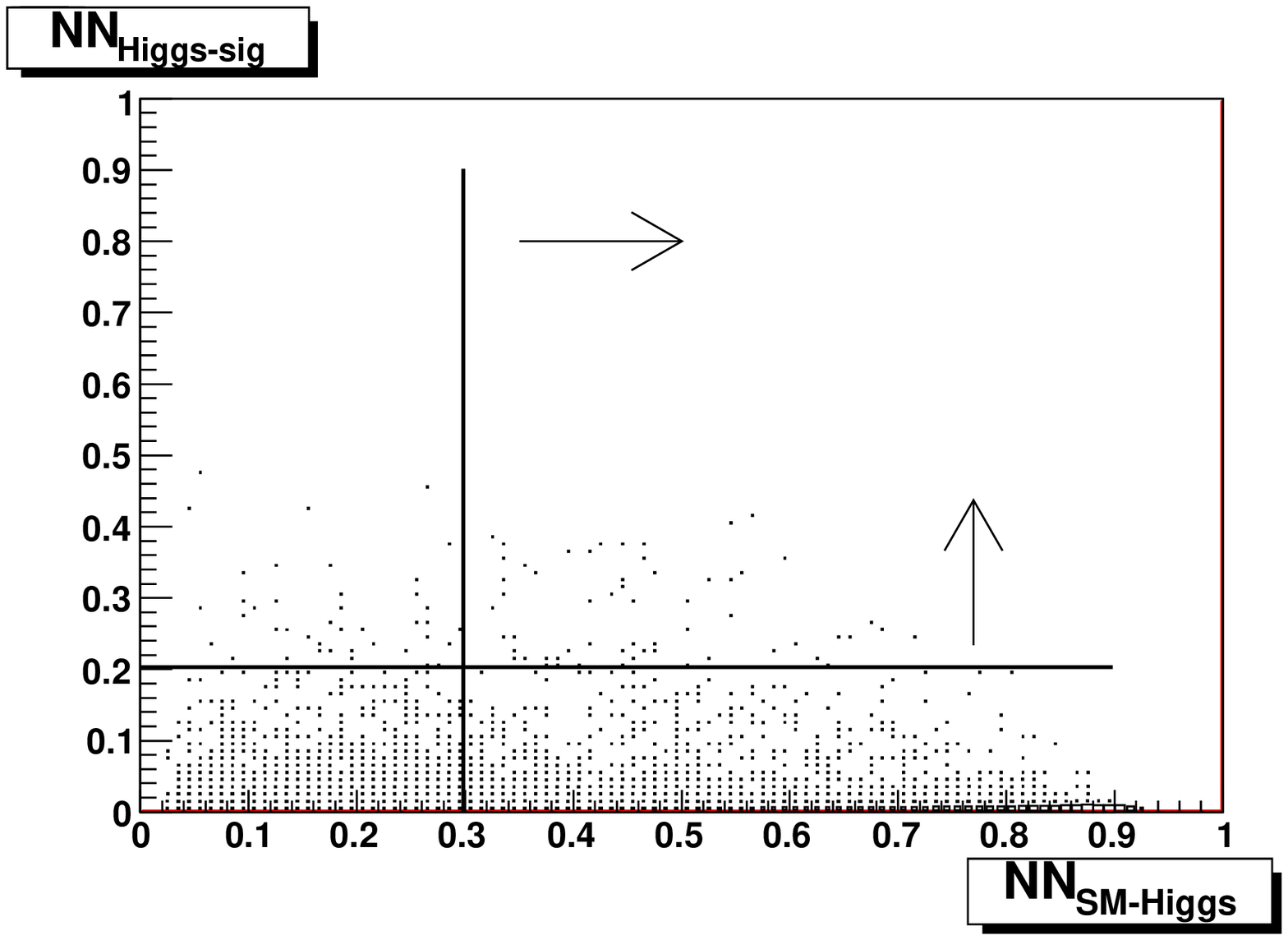}}
\subfloat[]{\includegraphics[scale=0.30]{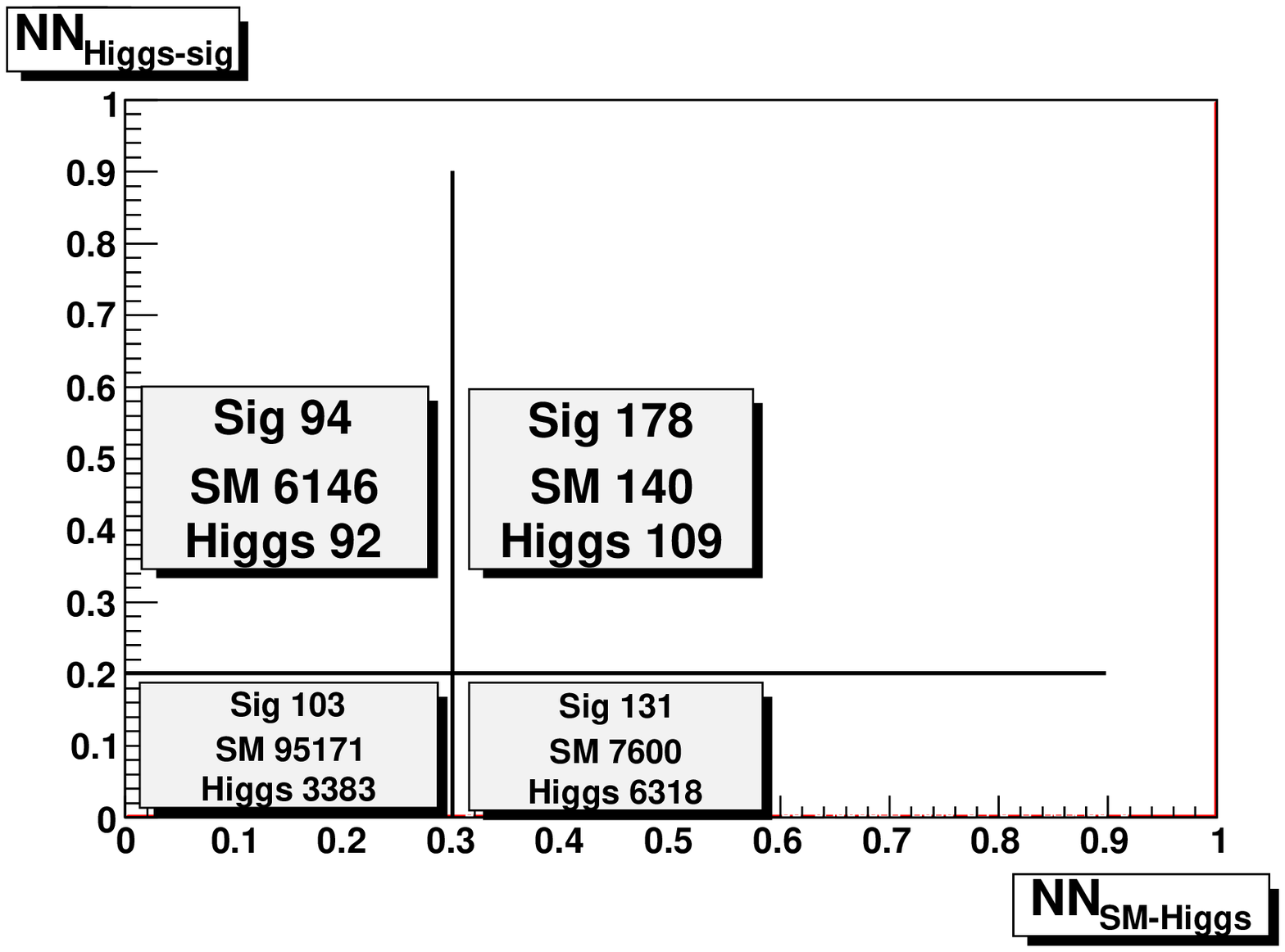}}
\end{center}
\caption{NN$_{Higgs-signal}$ versus NN$_{SM-Higgs}$ for (a) Signal, (b) Standard Model background, (c) Higgs background and (d) number of events for different NN regions}
\label{fig:n1n2}
\end{figure}

\begin{figure}[htbp]
\begin{center}
\subfloat[]{\includegraphics[scale=0.40]{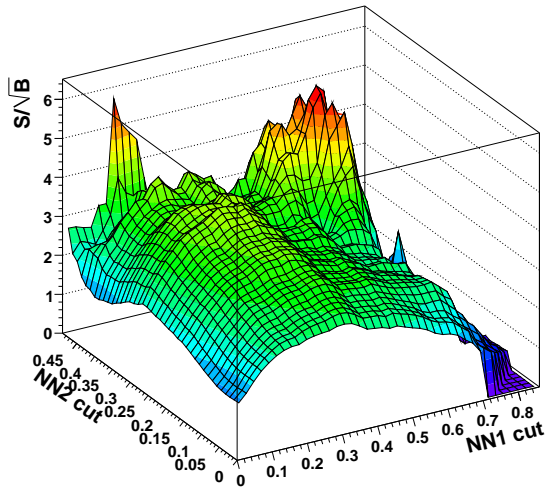}}
\end{center}
\caption{Neural Network Cut optimization for $c\bar{c}$ in the neutrino channel.}
\label{nnccuts}
\end{figure}

Figure~\ref{fig:n1n2} shows scatter plots of NN$_{Higgs-signal}$ versus NN$_{SM-Higgs}$ for the signal and backgrounds, and also shows the number of events in different regions of NN$_{Higgs-signal}$ and NN$_{SM-Higgs}$ parameter space.
The final event sample is required to have NN$_{Higgs-signal} > 0.2$ and NN$_{SM-Higgs} > 0.3$. This cut is optimised by using 
different values of NN$_{Higgs-signal}$ and NN$_{SM-Higgs}$ and choosing a point at which the signal cross-section error is minimised (or $S/\sqrt{B}$ ratio is maximal). The optimization is shown Figure~\ref{nnccuts}. After the final selections the sample includes 178 signal events with SM background of 140 events and Higgs background of 109 events. The signal efficiency is about 28\% and 
the purity is 42\%. The efficiency is calculated as the ratio between the final number of signal events and the number of signal events after the classification but before the pre-selection cuts.

%\begin{table}[h]
%\centering
%\begin{tabular}{|c|c|c|c|}
%\hline
%NN$_{Higgs-signal}$ $\downarrow$ & \multicolumn{3}{|c|}{S$/$$\sqrt{B}$} \\ \hline
%NN $>$ 0.4 & 0.8 & 5.4 & 2.7 \\ \hline
%NN $>$ 0.2 & 1.0 & $\textbf{7.6}$ & 6.0 \\ \hline
%No cut & 0.3 & 0.7 & 1.0 \\ \hline 
%NN$_{SM-Higgs}$ $\rightarrow$ & No cut & NN $>$ 0.3 & NN $>$ 0.6 \\ \hline
%\end{tabular}
%\caption{Optimization of neural net cuts using S/$\sqrt{B}$ values in the neutrino channel.}
%\label{tab:Nun1n2}
%\end{table}

\subsection{The hadronic channel}

In the hadronic channel, events are forced to have four reconstructed jets and as in the two-jet case the $y_{min}$ variable is chosen to
differentiate from backgrounds with different numbers of jets. In this channel the $y_{min}$ parameter is the minimum value of the y-parameter that leads to a 4-jet event configuration. For the signal, two of the jets should have di-jet invariant mass consistent with the Higgs boson and the other two having
di-jet mass consistent with the Z boson. 

Jet pairing is performed before kinematic fitting. For the 4-jet events we have 6 possible pairings of the jets and 3 possible associations of the 4 jets to the Z and H bosons. For the 6 possible pairings we calculate the invariant mass of each pair and compare with the masses of bosons. For each event we calculate 
\begin{equation}
d = (m_{ij}-m_{Z})^2 + (m_{kl}-m_{H})^2
\end{equation}
The pairing that minimises d is chosen and we find on average that jets 1 and 3 are associated to the Higgs and jets 2 and 4 are associated to the Z boson,
where the ranking is performed according to jet energy (jet 1 has the highest energy). 

Kinematic fitting~\cite{kinfit} is performed to improve identification of two jets from Higgs boson and two jets from Z-boson in order to reduce the combinatorial background. The kinematic fit minimizes a global chi square which is obtained from the difference between fitted and measured parameters of the kinematic variables, which are weighted by their covariance matrices. The following constraints are used for the analysis:
\begin{itemize}
\item $\displaystyle\sum_{i=0}^{3}P_i$ = 0 
\item sum of E = 250 GeV
\item invariant mass difference of two jet pairs is equal to difference between Higgs and Z boson masses
\end{itemize}

\begin{figure}[htbp]
\begin{center}
\subfloat[]{\includegraphics[scale=0.30]{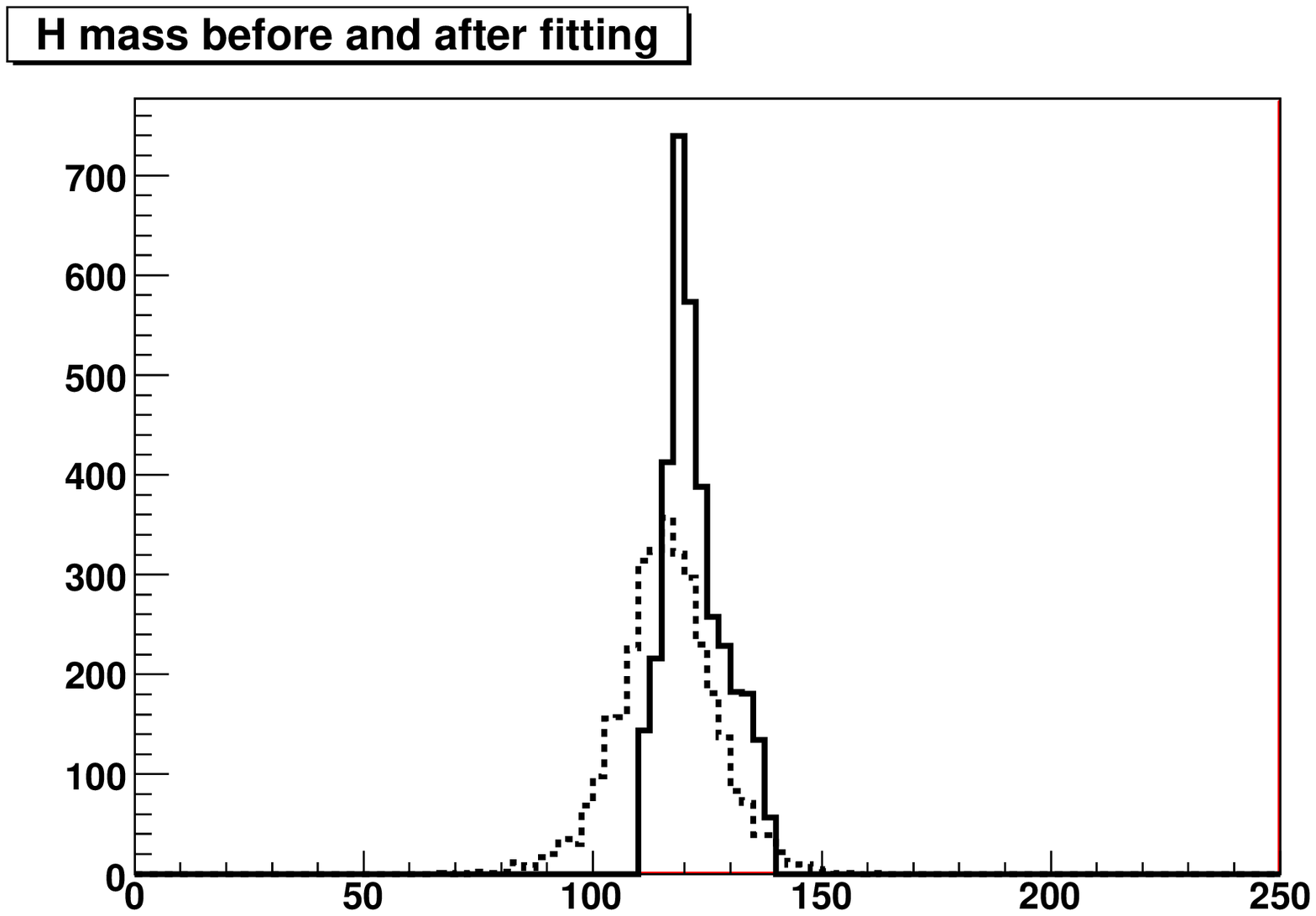}}
\subfloat[]{\includegraphics[scale=0.30]{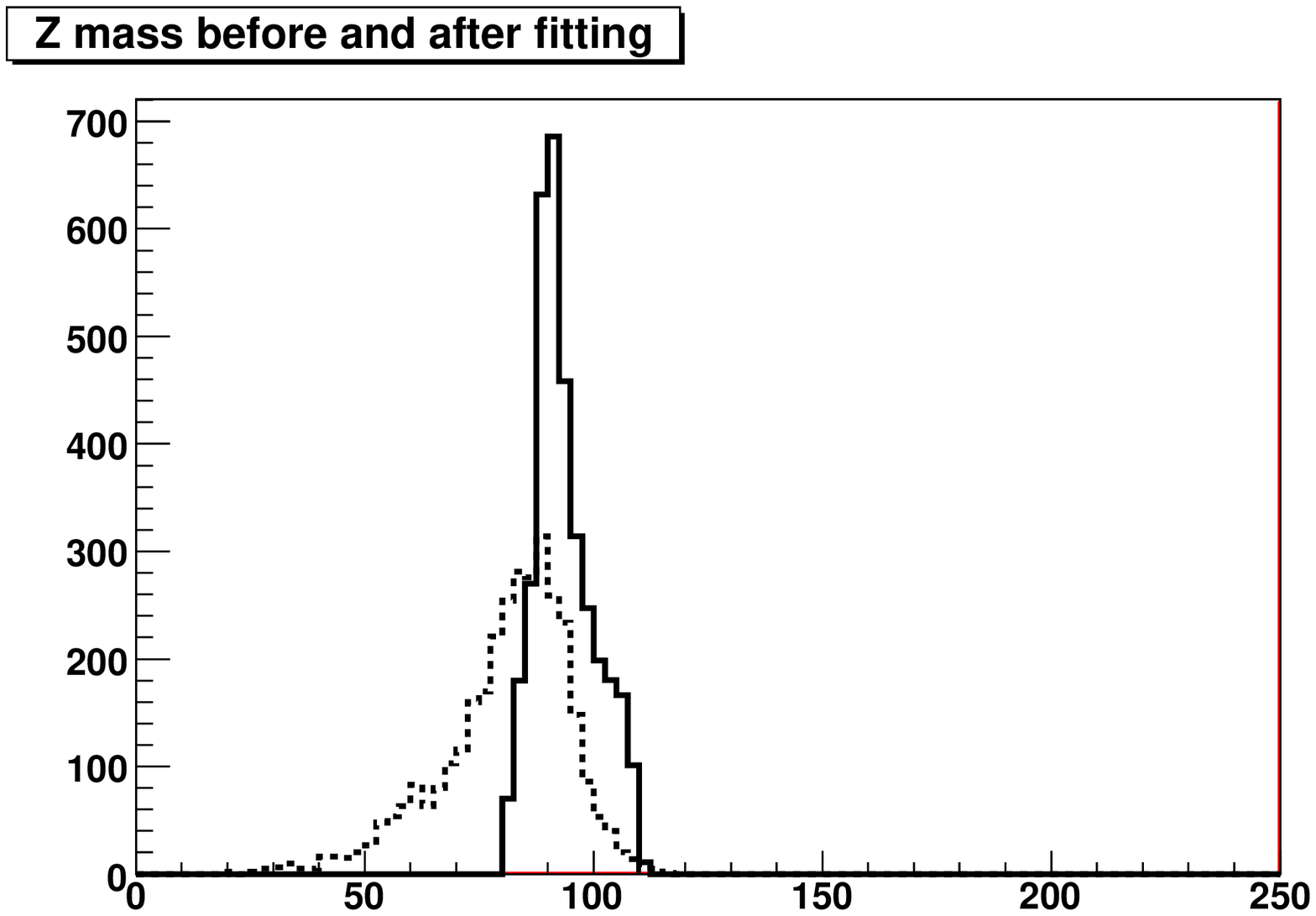}}
\end{center}
\caption{Higgs (a) and Z (b) invariant masses before (dashed line) and after (solid line) kinematic fitting.}
\label{fig:had}
\end{figure}

The main backgrounds for this channel are WW and ZZ pairs where the all bosons decay to hadrons. Figure~\ref{fig:had} shows the Higgs and Z invariant masses before and after kinematic fitting.

\begin{figure}[htbp]
\begin{center}
\subfloat[]{\includegraphics[scale=0.20]{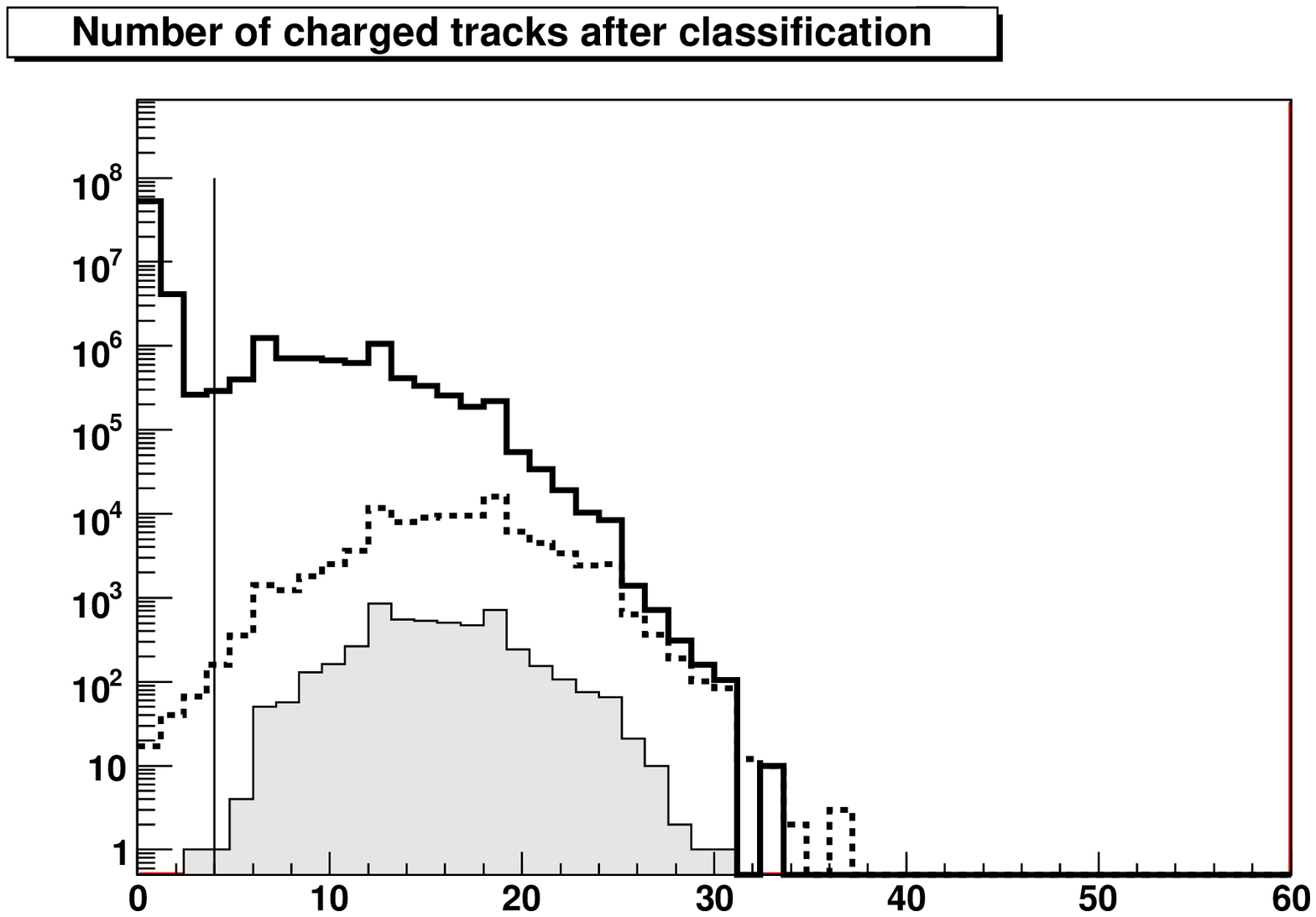}}
\subfloat[]{\includegraphics[scale=0.20]{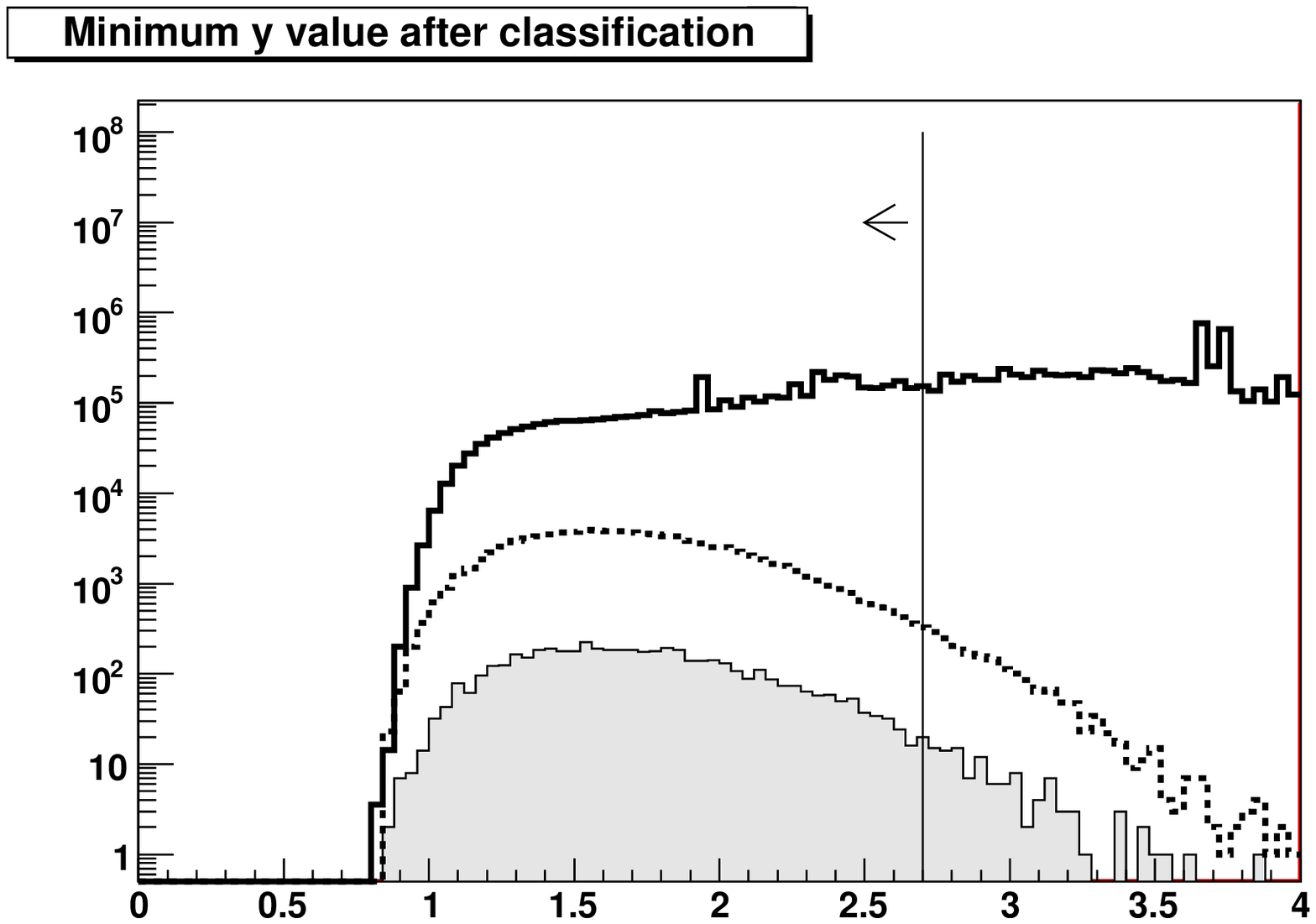}} \\
\subfloat[]{\includegraphics[scale=0.20]{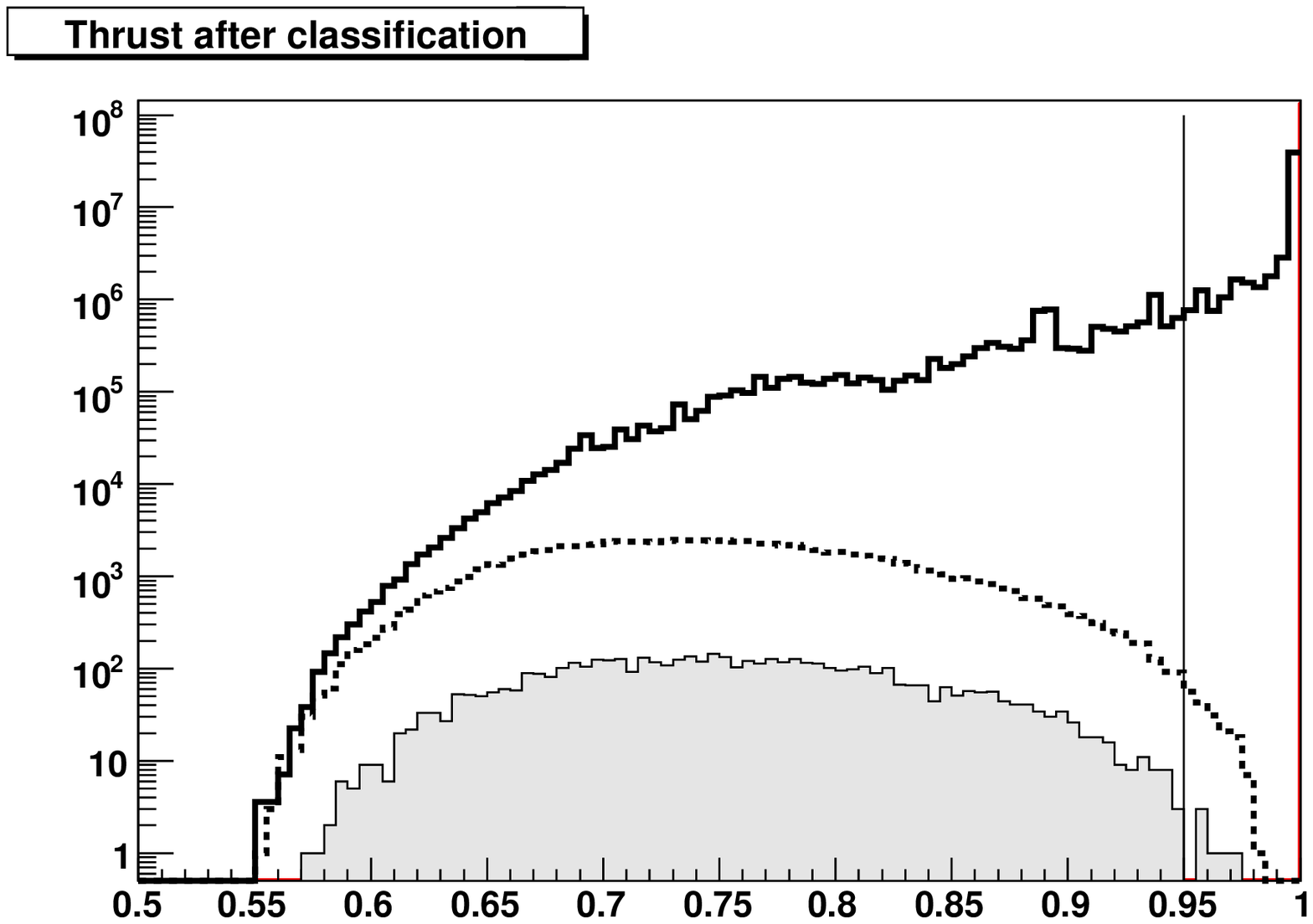}}
\subfloat[]{\includegraphics[scale=0.20]{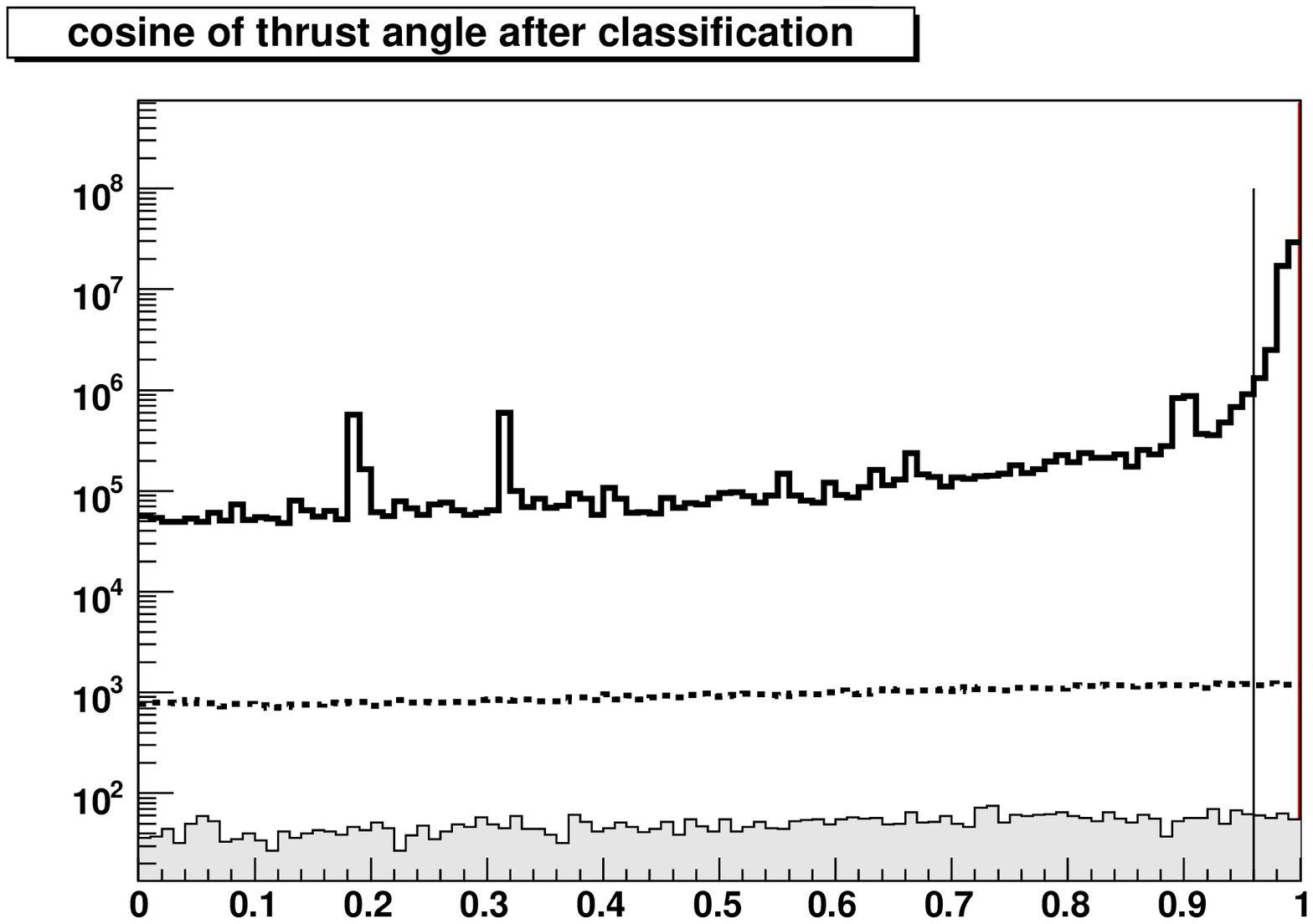}} \\
\subfloat[]{\includegraphics[scale=0.20]{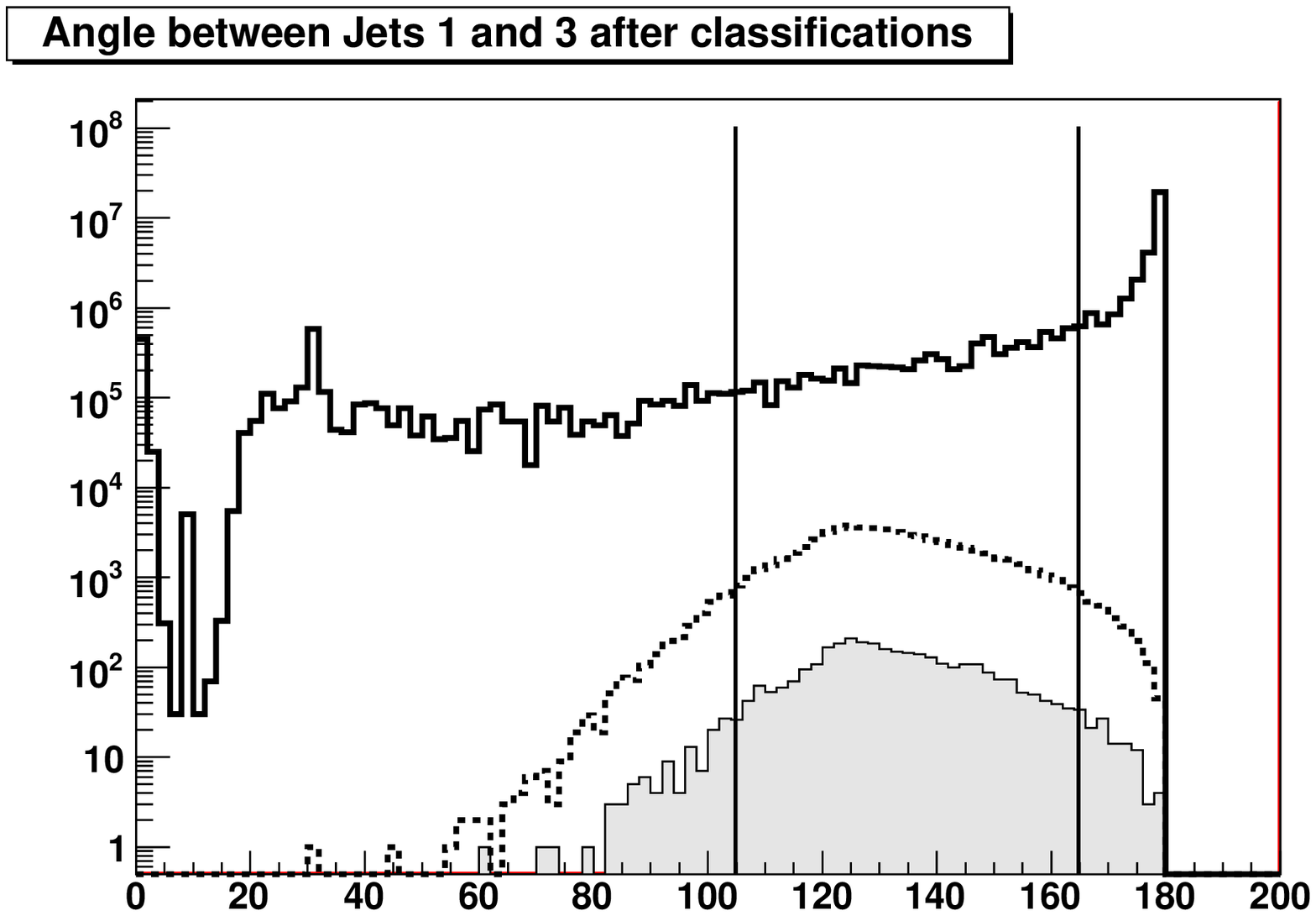}}
\subfloat[]{\includegraphics[scale=0.20]{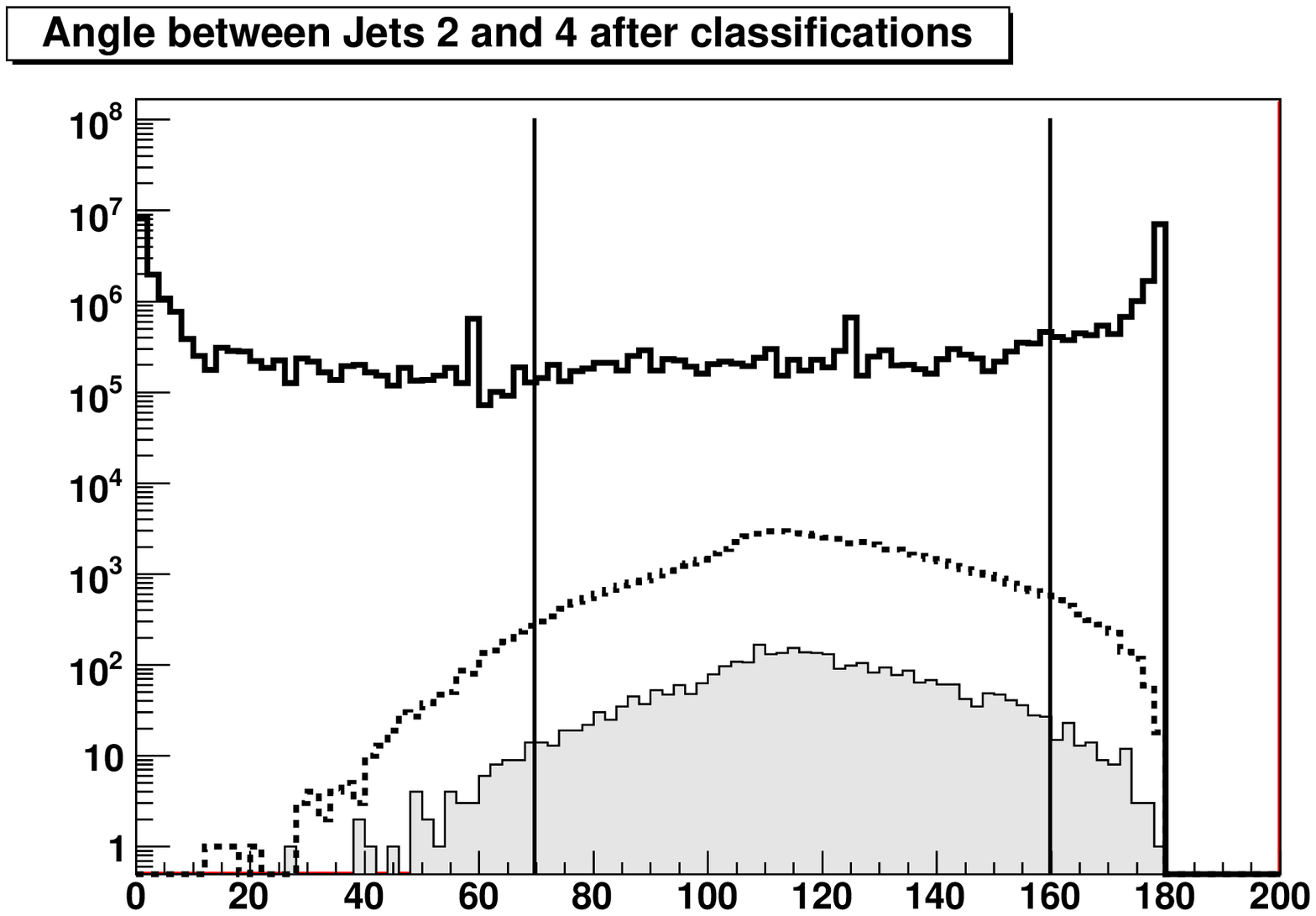}} \\
\subfloat[]{\includegraphics[scale=0.20]{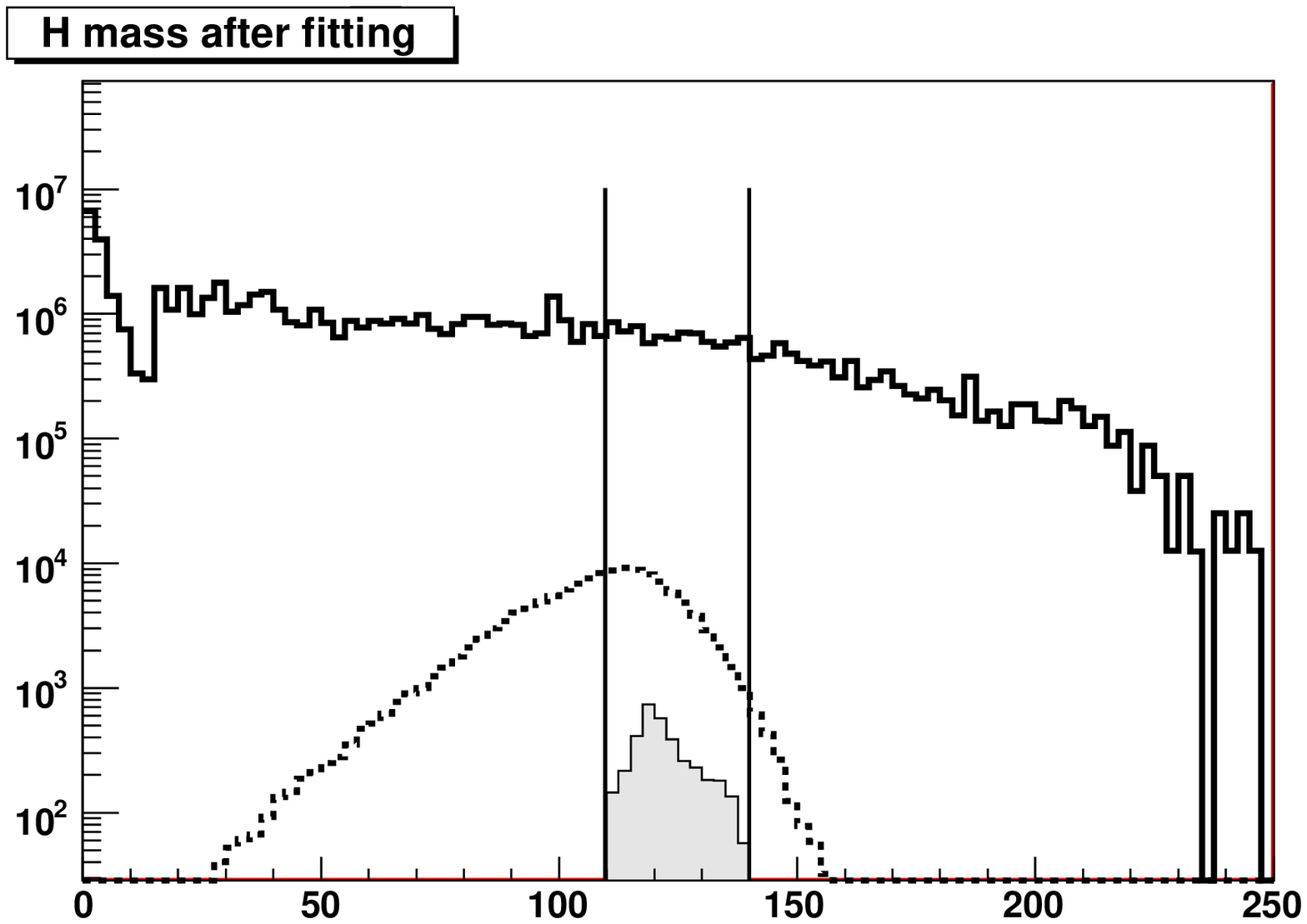}}
\subfloat[]{\includegraphics[scale=0.20]{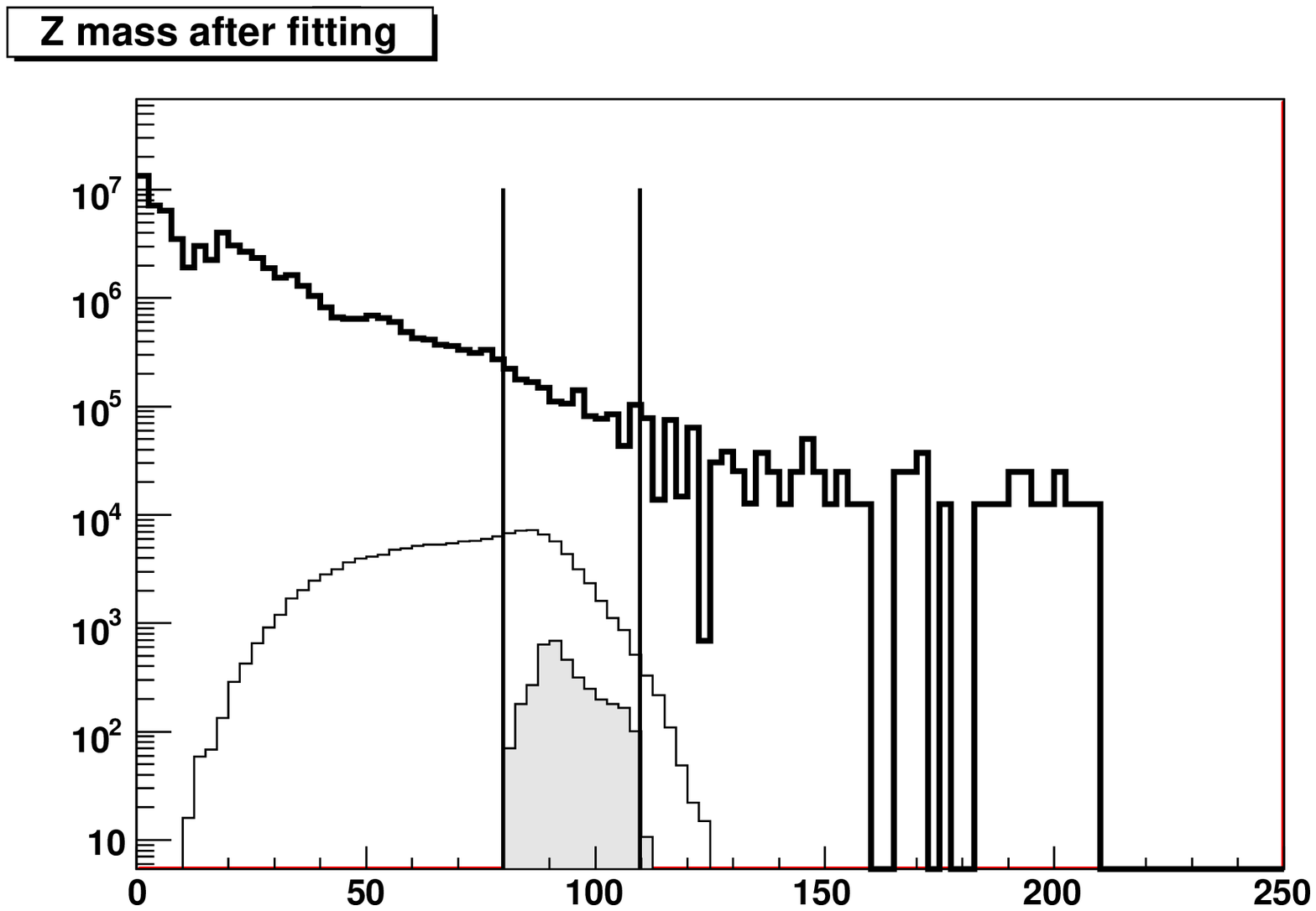}} \\
\subfloat[]{\includegraphics[scale=0.20]{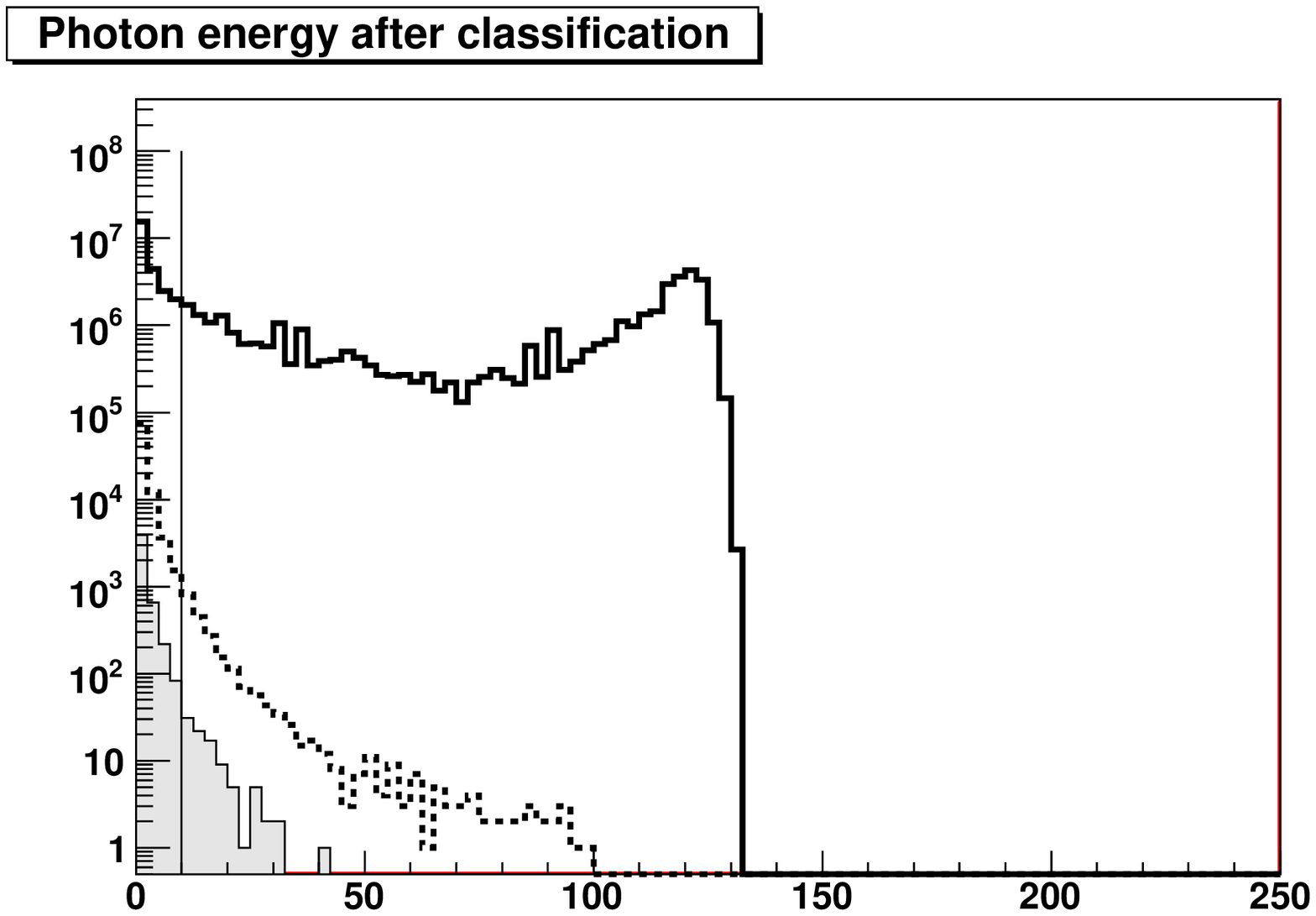}}
\end{center}
\caption{Distribution of pre-selection kinematic variables. Solid curves are SM background, dashed curves are Higgs background
sample and filled histograms are the signal. See text for discription of variables.}
\label{fig:had2}
\end{figure}

For the preselection, we use kinematic cuts similar to the cuts for the neutrino channel. Figure~\ref{fig:had2} shows distributions of variables used for pre-selection for the signal and backgrounds after classification. The variables are as defined in Section 4.1.

Selections used for further reduction of the background are presented in Table~\ref{tab:HZ2}.
For the cuts, the jets were ordered in energy and the invariant mass and angle cuts were employed after the kinematic fitting.
Table~\ref{tab:4Nuevts} shows the number of events before and after pre-selection cuts.

\begin{figure}[htbp]
\begin{center}
\subfloat{\includegraphics[scale=0.30]{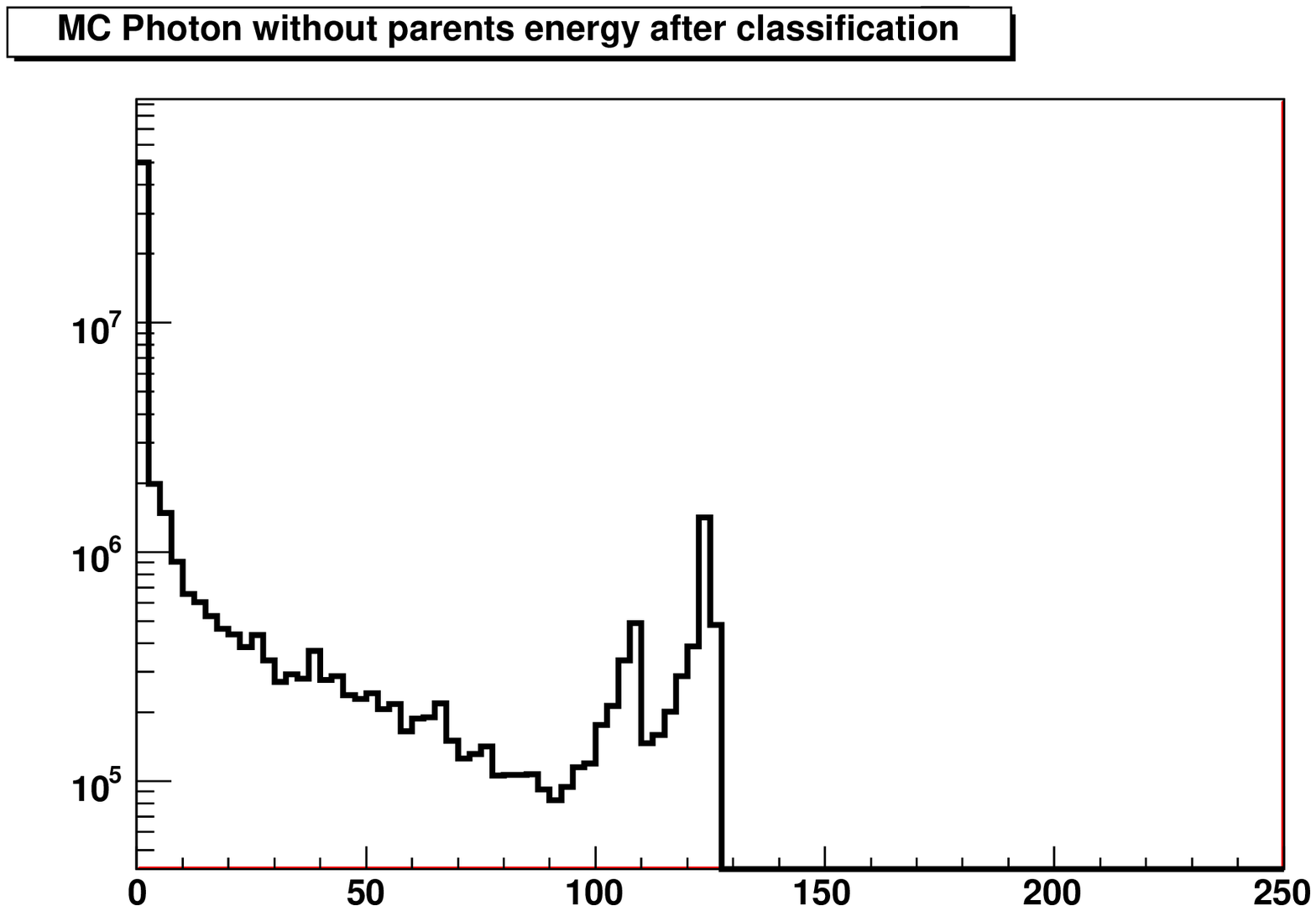}}
\subfloat{\includegraphics[scale=0.30]{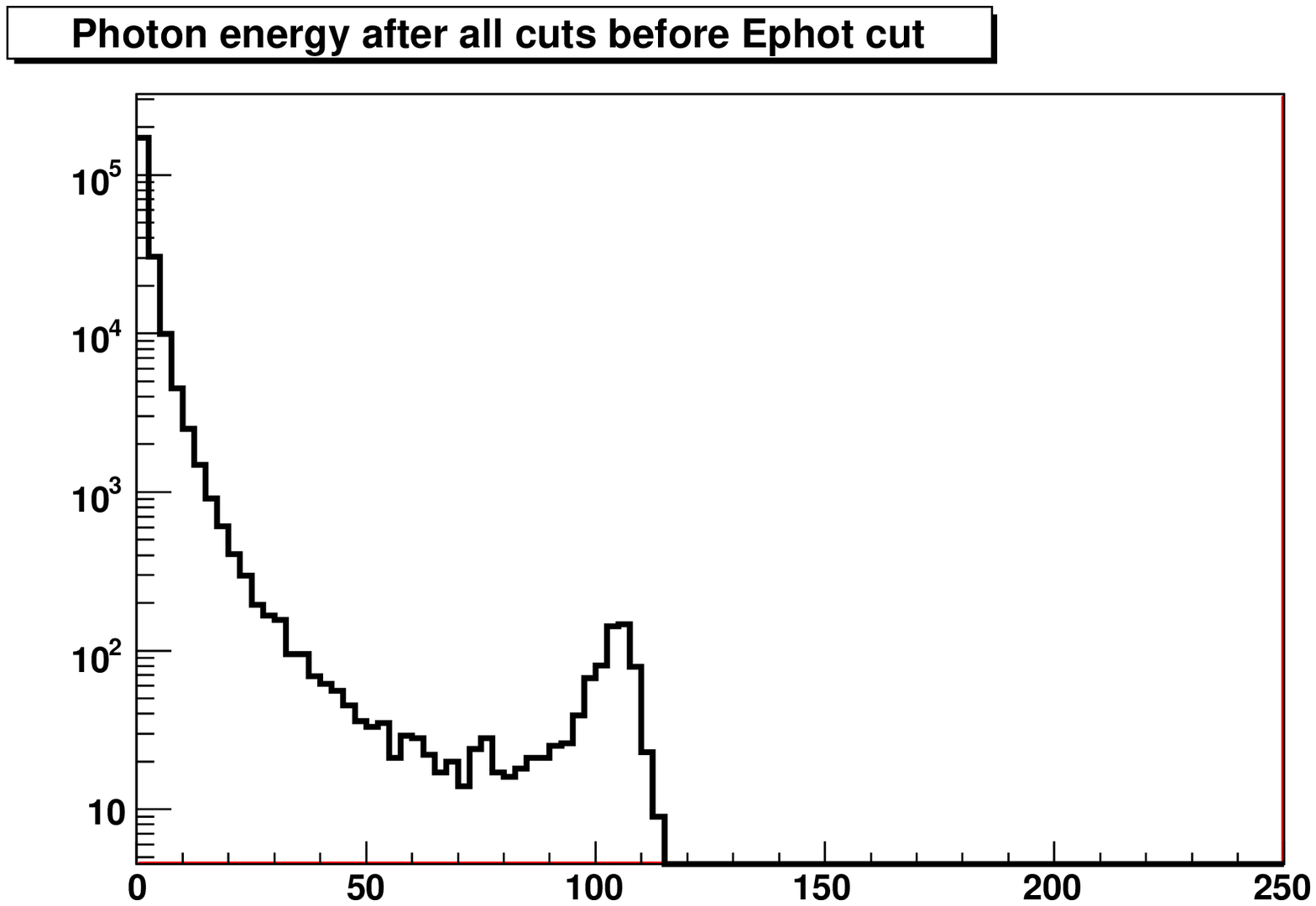}}
\end{center}
\caption{(a) Energy of MC photon for the Standard Model background after channel classification; (b) Energy of reconstructed photon for the Standard Model background after channel classification and kinematic selection cuts.}
\label{fig:MCphot}
\end{figure}

For a 250 GeV centre-of-mass scenario, the energy of the photon resulting from radiative returns to the Z-peak is expected to be at around 108.6 GeV. Figure~\ref{fig:had2}(i) shows the distribution of the SM background reconstructed photon energy after channel classification but before any kinematic selections. There is a peak seen at about 125 GeV which does not correspond to the radiative return to Z-peak. The main contributions to this peak are events where a colliding e$^{+}$ or e$^{-}$ loses most of its energy by ISR. Figure~\ref{fig:MCphot}(a) shows the energy of all SM background photons at the generator level for the hadronic channel after the channel classification. This distribution has two peaks at about 108 GeV and 125 GeV, which confirms that these photons originate from both Z radiative returns and direct ISR. The first peak corresponds to the radiative return to the Z-peak and the second peak corresponds to the peak observed in Figure~\ref{fig:had2}(i). Figure~\ref{fig:MCphot}(b) shows the SM reconstructed photon energy after channel classification but here all kinematic selections (1-8) are applied. Here we notice that we now have a peak at 108 GeV and no peak at around 125 GeV. We conclude that the previous selections are successful in removing direct ISR photons in the other peak.  

\begin{table}[h]
\centering
\begin{tabular}{lllll}
\hline Cuts & && selection & value  \\
\hline
1. & && number of charged tracks per jet & $>$ 4\\
2. & && $-\log(y_{min}$) & $<$ 2.7 \\
3. & && thrust & $< 0.95$ \\
4. & && $ \cos(\theta_{thrust})$ & $< 0.96$ \\
5. & $105^{\circ}$ &$<$ & angle between jet 1 and 3 &$< 165^{\circ}$ \\
6. & $70^{\circ}$ &$<$ & angle between jet 2 and 4 &$< 160^{\circ}$ \\
7. & 110 GeV &$<$ & invariant mass of Higgs candidate after fit &$<$ 140 GeV \\
8. & 80 GeV &$<$ & invariant mass of Z candidate after fit &$<$ 110 GeV \\
9. & && Highest reconstructed photon energy & $<$ 10 GeV \\
\hline
\end{tabular}
\caption{Selections for the four-jet analysis.}
\label{tab:HZ2}
\end{table}

\begin{table}[htpb]
\centering
\begin{tabular}{|c|c|c|c|}
\hline
Cuts & SM background & Higgs background & Signal \\ \hline
(i)Before Classification  & 9275594683 & 51513 & 1931 \\ \hline
(0)After Classification & 39398366 & 41016 & 1837 \\ \hline 
(1) & 18601753 & 30125 & 1143 \\ \hline
(2) & 13921271 & 29478 & 1101 \\ \hline
(3) & 8737017 & 27065 & 1047\\ \hline
(4) & 7943851 & 26322 & 1017 \\ \hline
(5) & 5871237 & 26001 & 979 \\ \hline
(6) & 4898312 & 25687 & 978 \\ \hline
(7) & 1917231 & 22533 & 966 \\ \hline
(8) & 1561432 & 21877 & 963 \\ \hline
(9) & 967312 & 15687 & 947 \\ \hline
\end{tabular}
\caption{Number of $c\bar{c}$ events before channel classification, after channel classification and after pre-selection cuts in the hadronic mode.}
\label{tab:4Nuevts}
\end{table}

\begin{figure}[htbp]
\begin{center}
\subfloat[]{\includegraphics[scale=0.25]{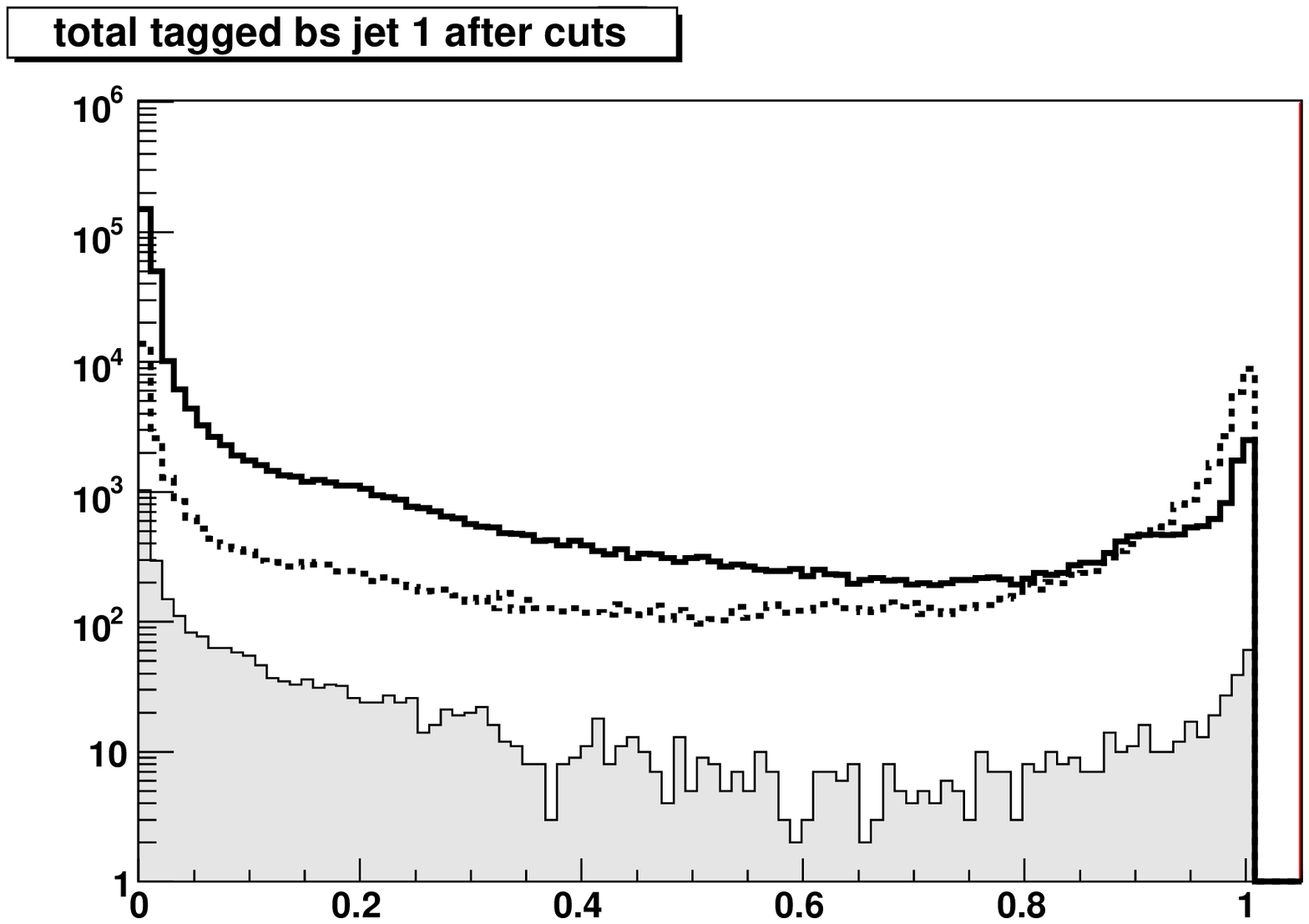}}
\subfloat[]{\includegraphics[scale=0.25]{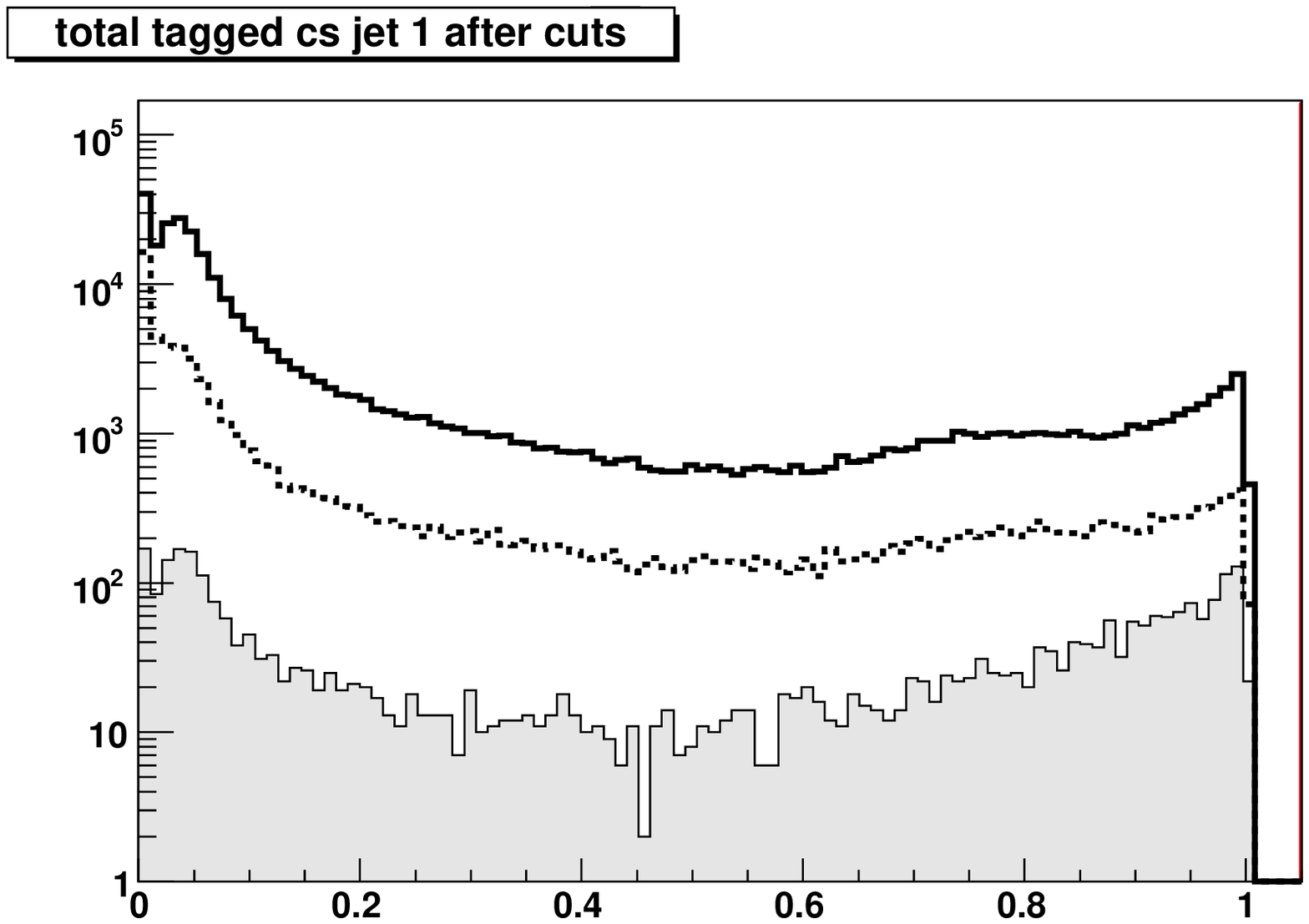}} \\
\subfloat[]{\includegraphics[scale=0.25]{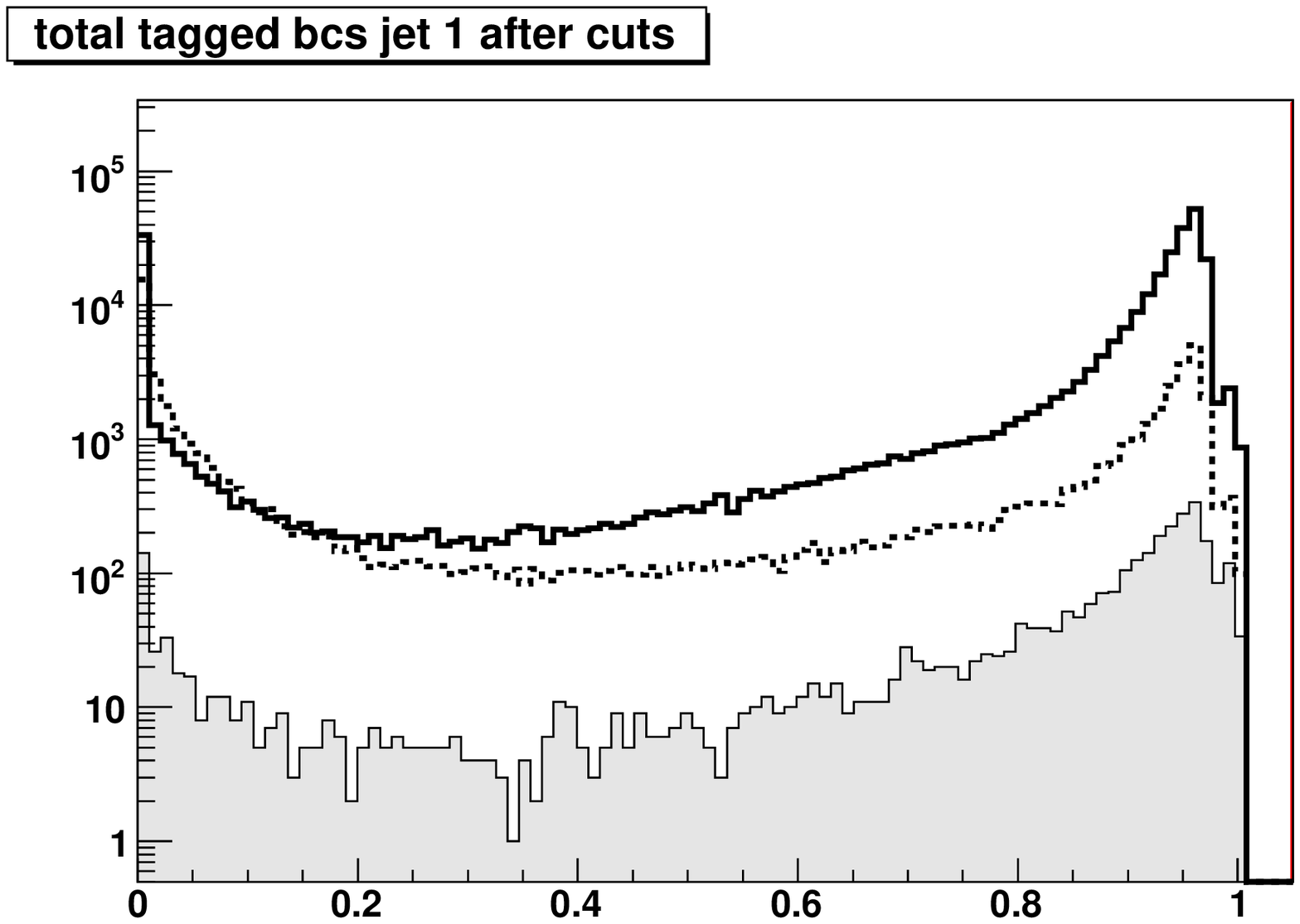}} \\
\subfloat[]{\includegraphics[scale=0.25]{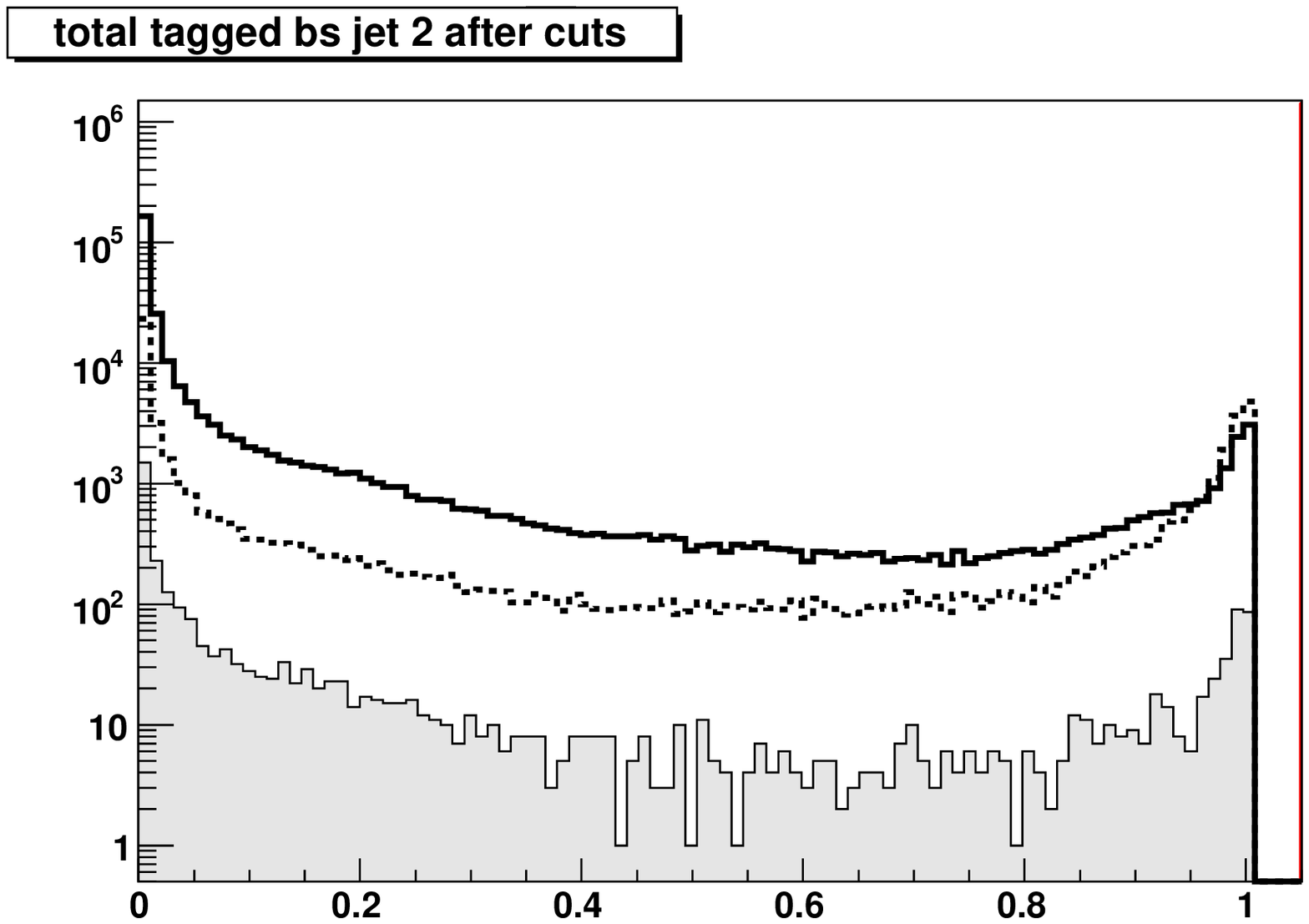}}
\subfloat[]{\includegraphics[scale=0.25]{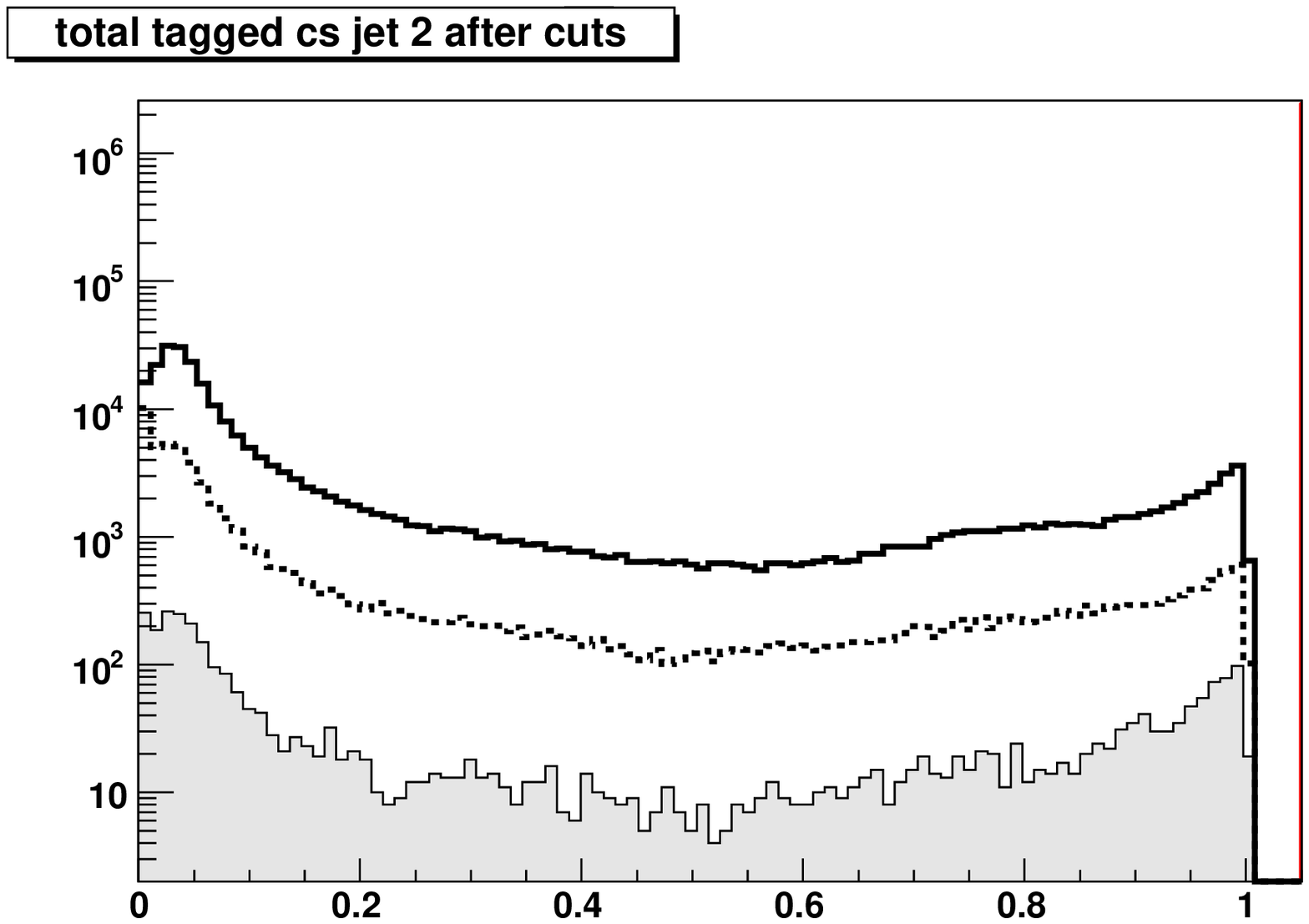}} \\
\subfloat[]{\includegraphics[scale=0.25]{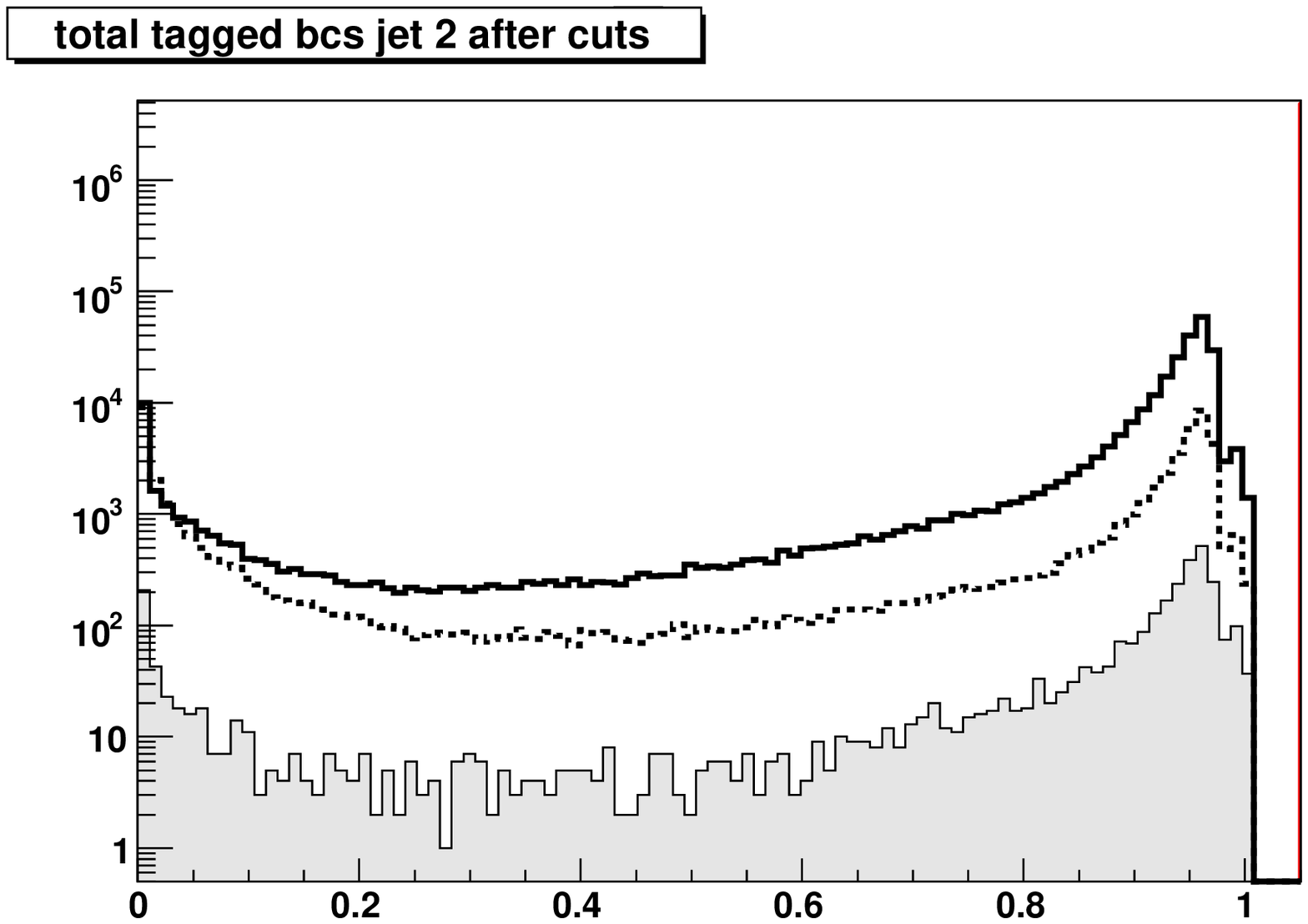}}
\end{center}
\caption{Hadronic channel: (a) first jet b-tag; (b) first jet c-tag; (c) first jet c-tag with b background only
(d) second jet b-tag; (e) second jet c-tag; (f) second jet c-tag with b background only. Solid curves are SM 
background, dashed curves are Higgs background sample and filled histograms are the signal.}
\label{fig:ccqqNN}
\end{figure}

\begin{figure}[htbp]
\begin{center}
\subfloat[]{\includegraphics[scale=0.30]{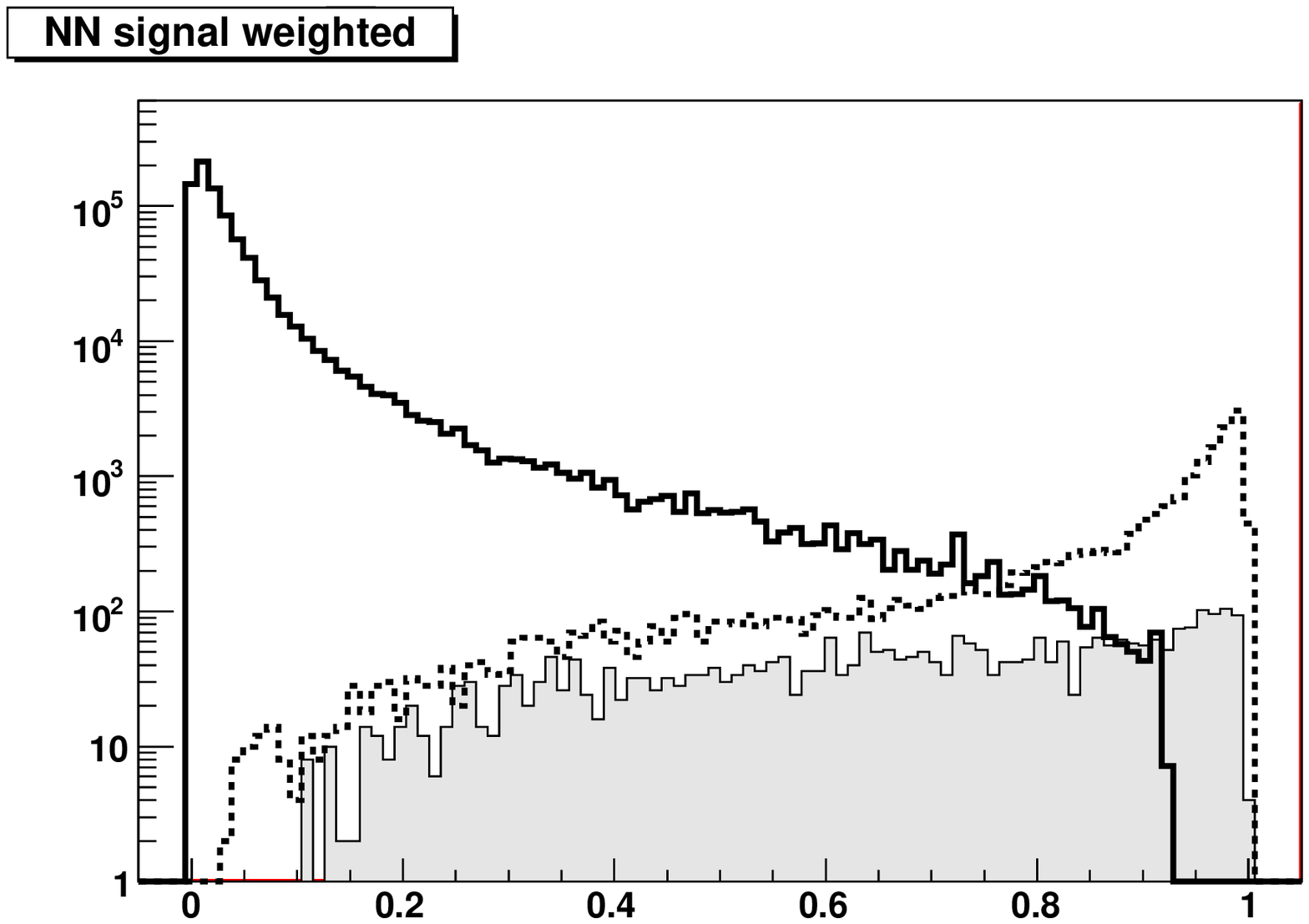}}
\subfloat[]{\includegraphics[scale=0.30]{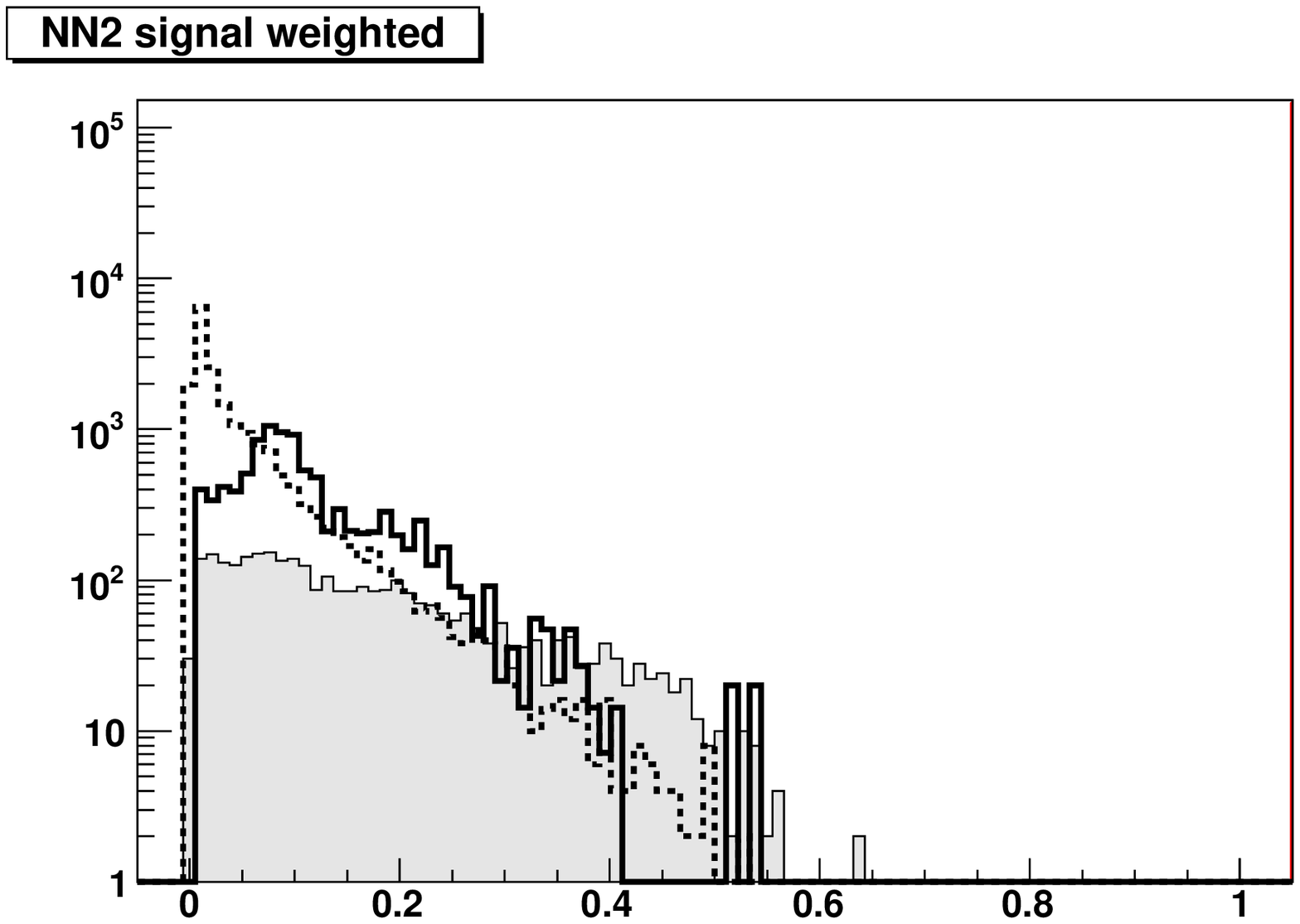}}
\end{center}
\caption{Hadronic channel: (a) NN$_{SM-Higgs}$ output; (b) NN$_{Higgs-signal}$ output. Solid curves are SM
background, dashed curves are inclusive Higgs sample and filled histograms are the signals.}
\label{fig:4nns}
\end{figure}

Similarly to the neutrino channel the variables above and also the three flavour tag outputs for all jets, are used in a neural network based selection employing the FANN package. Figure~\ref{fig:ccqqNN} shows the distribution of the three LCFI flavour tags, `b-tag', `c-tag' and  `c-tag with b background only' for the first and second leading jets. The first NN is trained to distinguish the SM background from the inclusive Higgs sample producing the NN$_{SM-Higgs}$ output, and, the second NN is trained to distinguish the signal from the Higgs background sample producing the NN$_{Higgs-signal}$ output. The outputs of the trained NNs are shown in Figure~\ref{fig:4nns}. 

\begin{figure}[htbp]
\begin{center}
\subfloat[]{\includegraphics[scale=0.30]{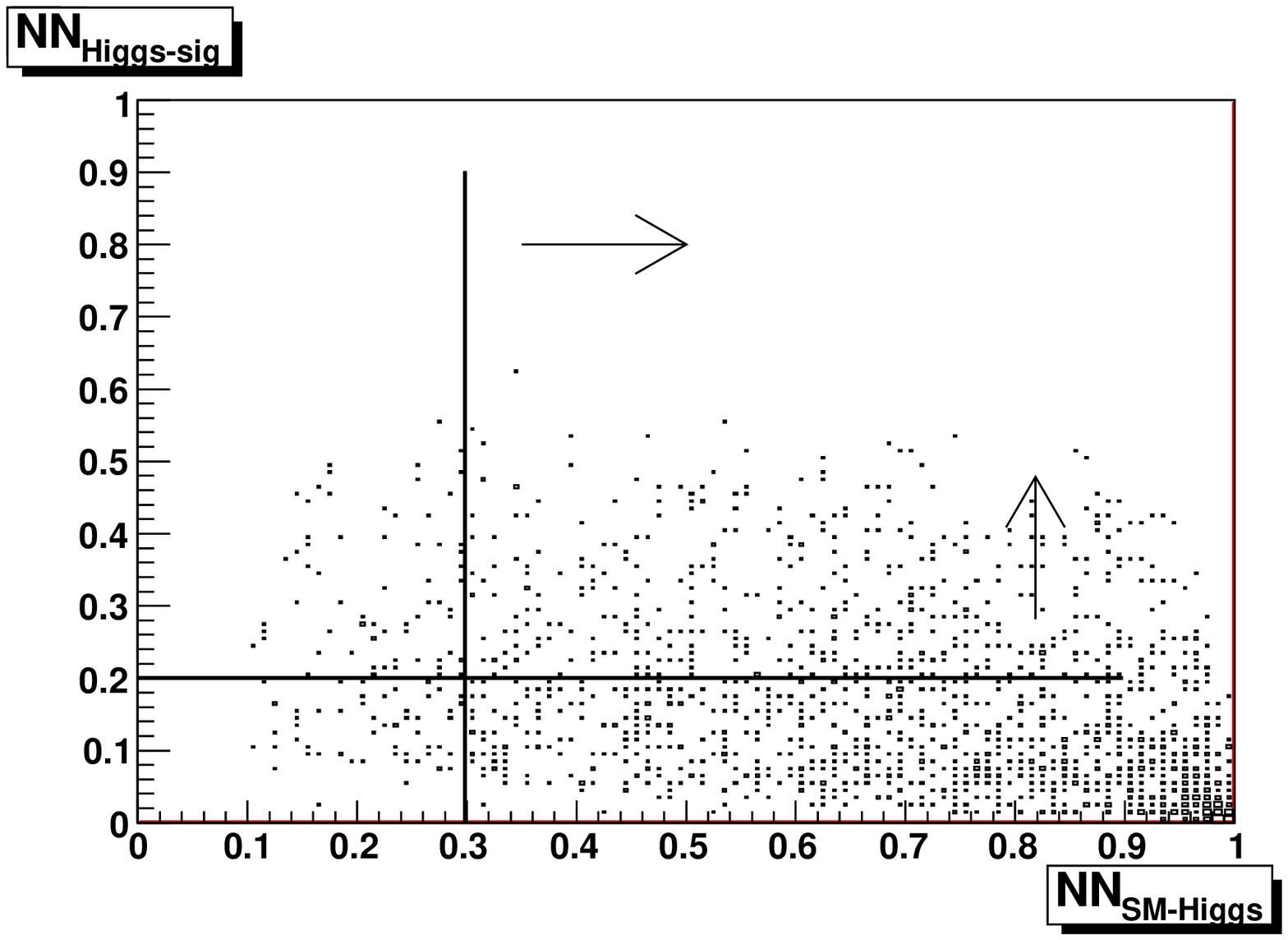}}
\subfloat[]{\includegraphics[scale=0.30]{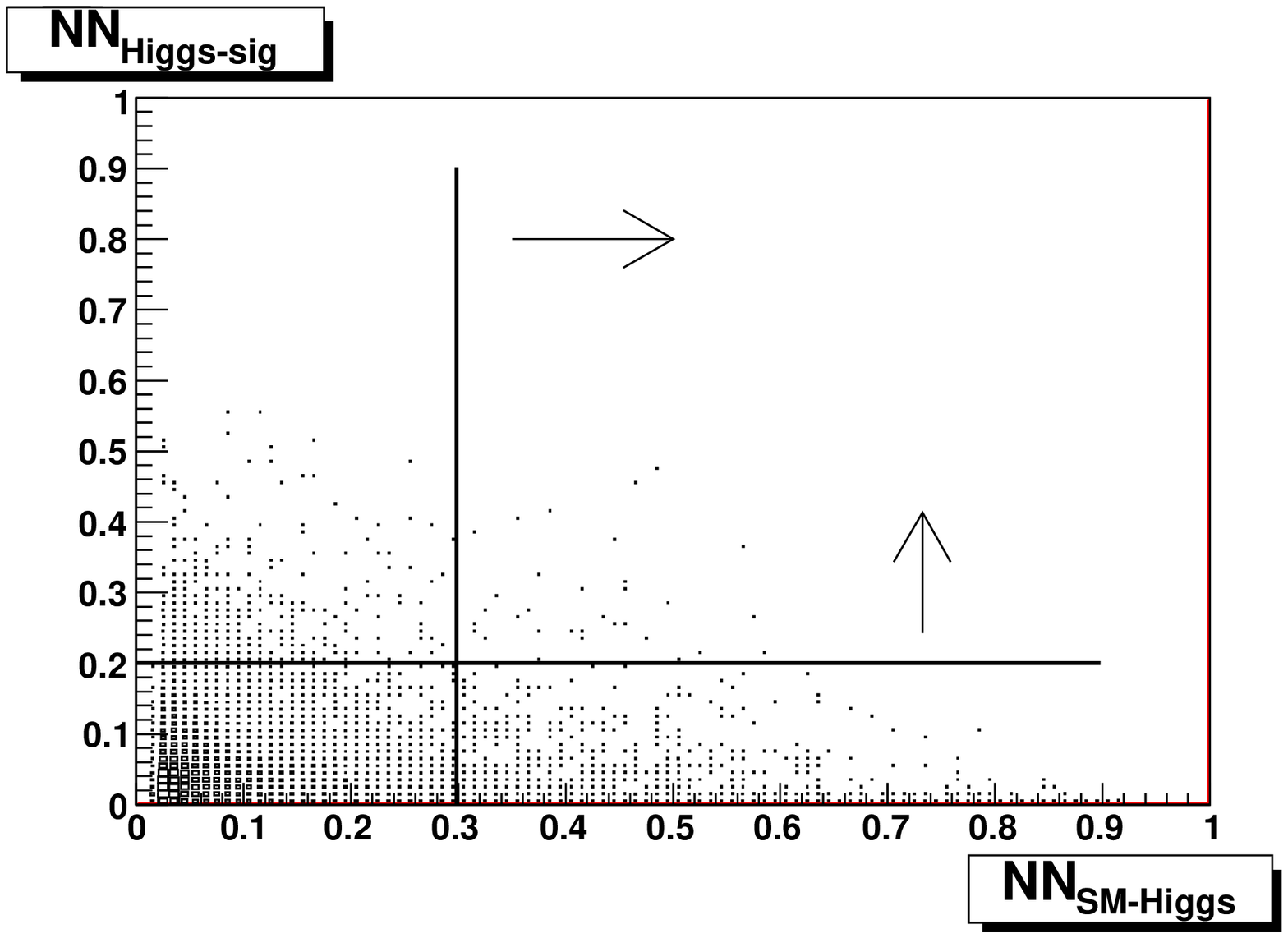}} \\
\subfloat[]{\includegraphics[scale=0.30]{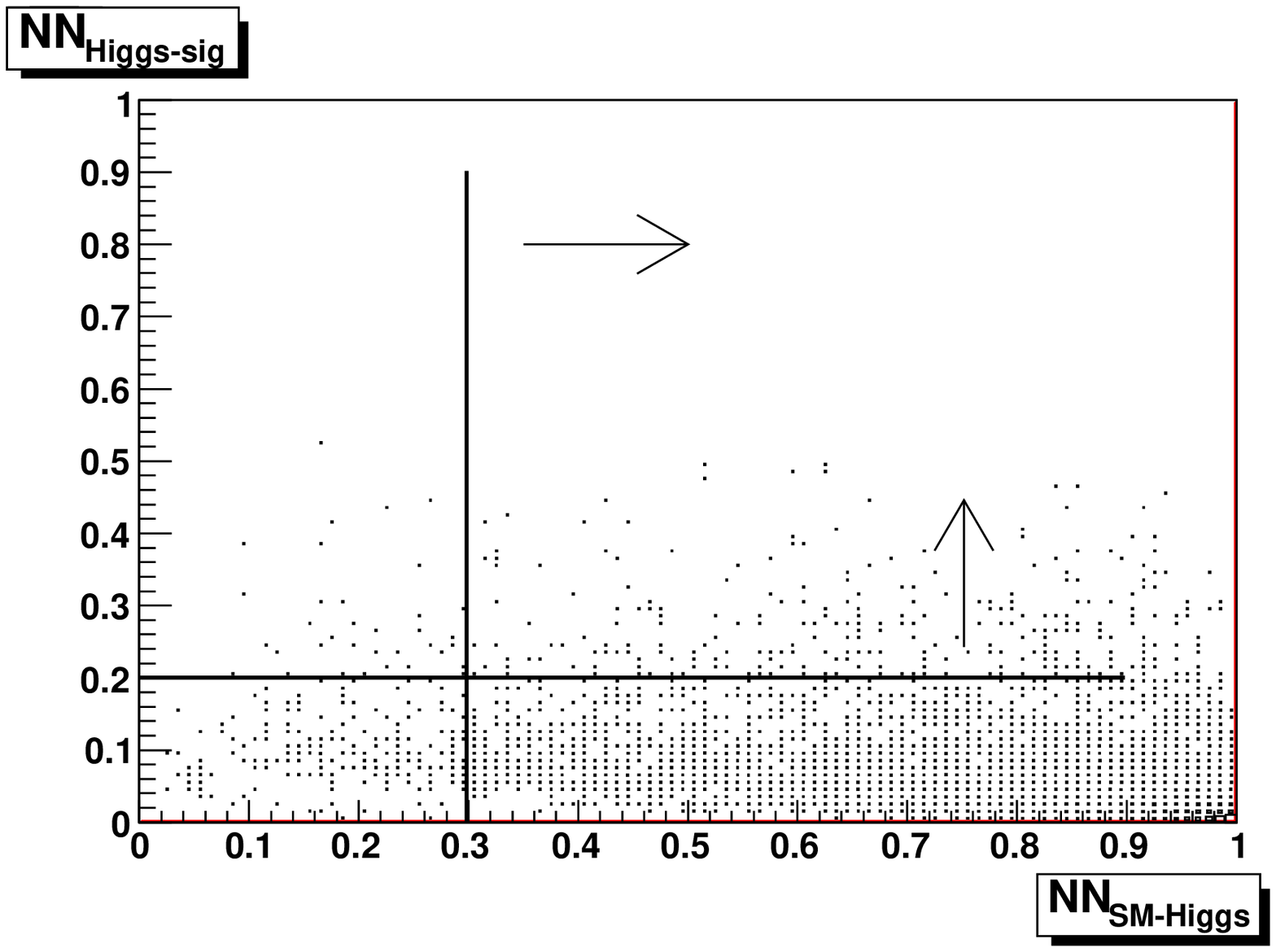}}
\subfloat[]{\includegraphics[scale=0.30]{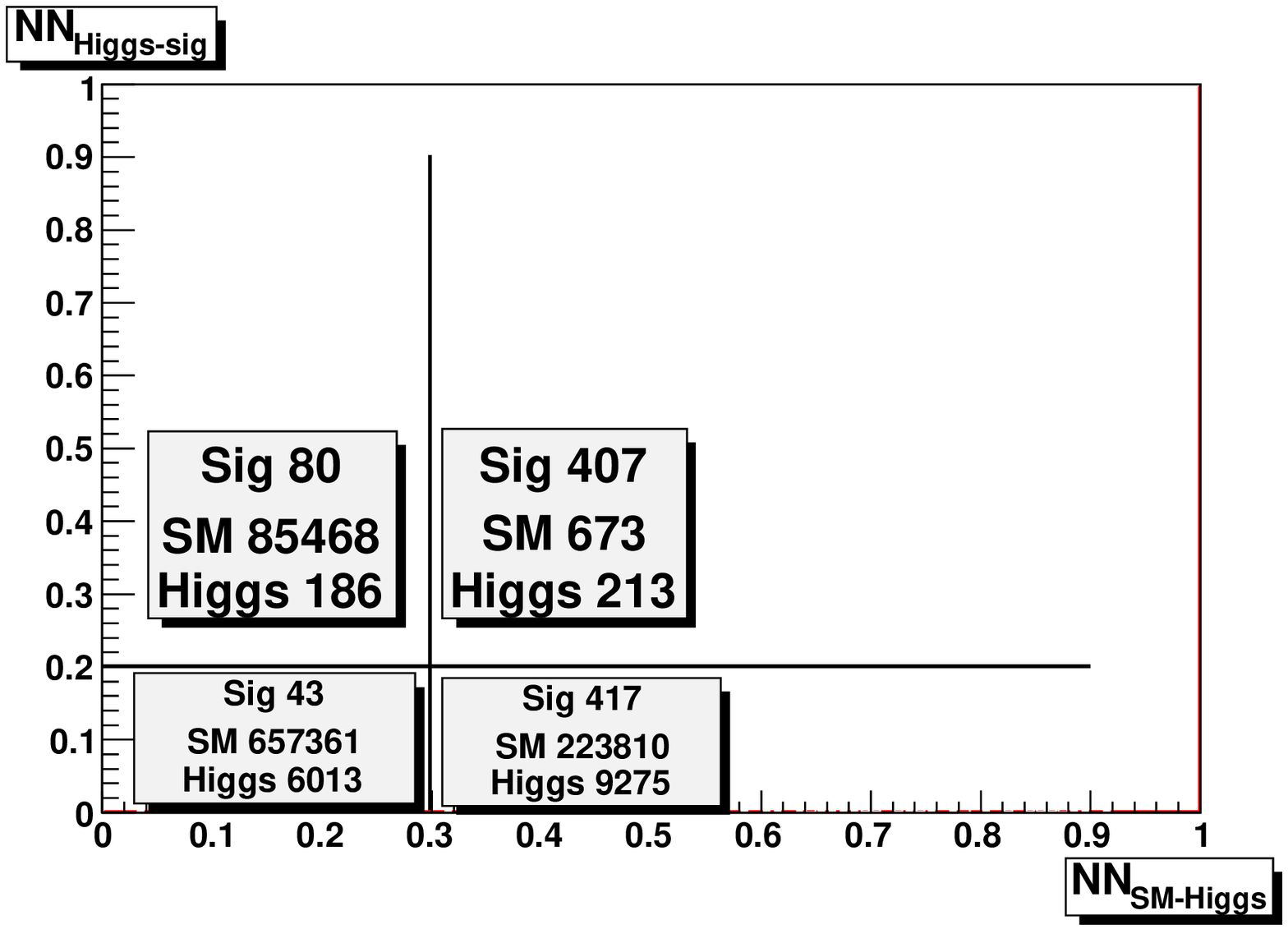}}
\end{center}
\caption{NN$_{Higgs-signal}$ versus NN$_{SM-Higgs}$ for (a) Signal, (b) Standard Model background, (c) Higgs background and (d) number of events for different NN regions.}
\label{fig:4n1n2}
\end{figure}

\begin{figure}[htbp]
\begin{center}
\subfloat[]{\includegraphics[scale=0.40]{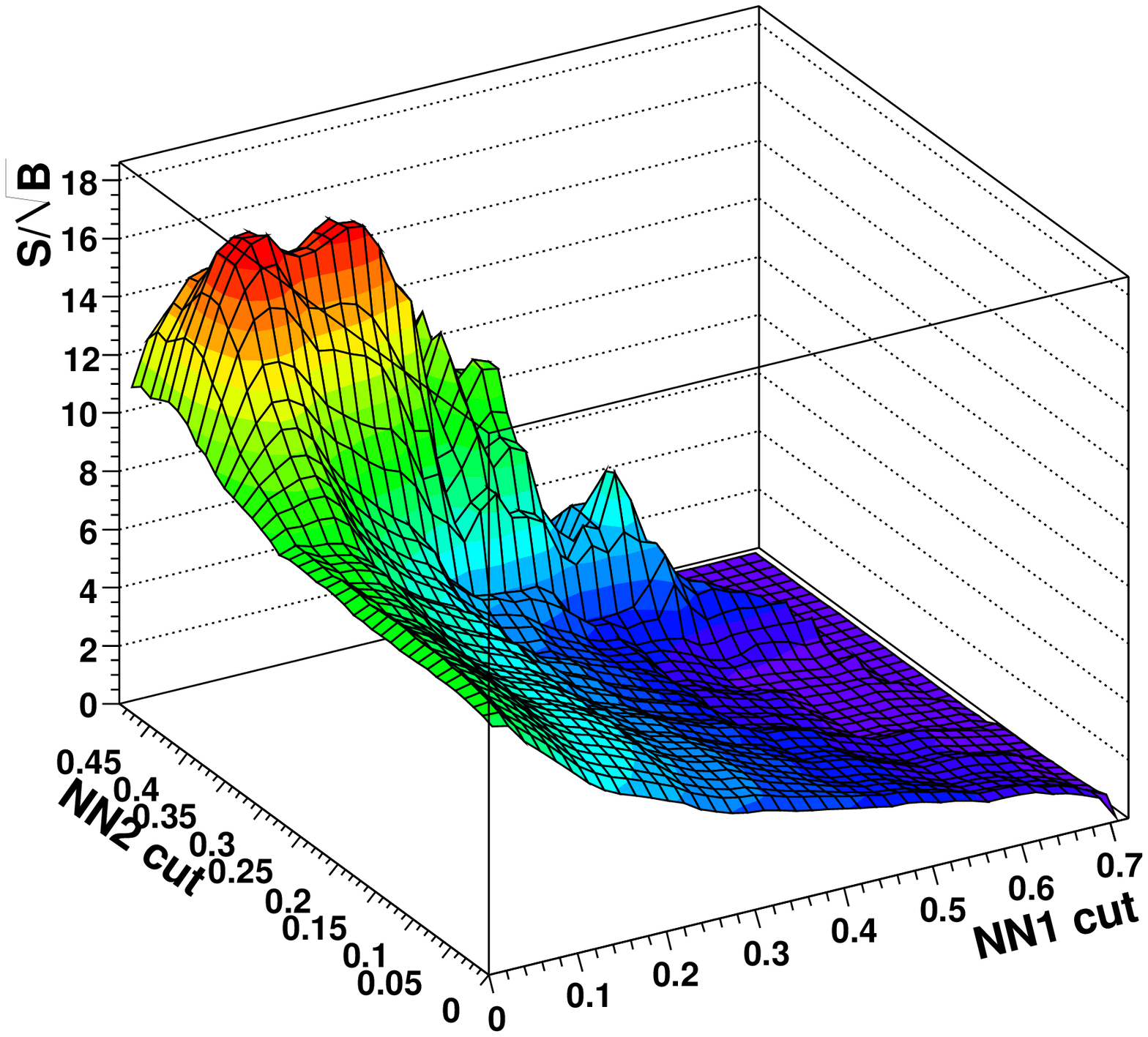}}
\end{center}
\caption{Neural Network Cut optimization for $c\bar{c}$ in the hadronic channel.}
\label{4nnccuts}
\end{figure}

Figure~\ref{fig:4n1n2} shows scatter plots of NN$_{Higgs-signal}$ versus NN$_{SM-Higgs}$ for the signal and backgrounds, and also shows the number of events in different regions of NN$_{Higgs-signal}$ and NN$_{SM-Higgs}$. The final signal sample is required to have NN$_{Higgs-signal} > 0.2$ and NN$_{SM-Higgs} > 0.3$. Optimisation of the NN cuts used the same method as in the neutrino channel. The optimization is summarized in Figure~\ref{4nnccuts}. After the final selections the total number of events sample includes 407 signal events with SM background of 673 events and Higgs background of 213 events. The signal efficiency (calculated as in the neutrino channel) is about 22\% and the purity is 32\%.

%\begin{table}[h]
%\centering
%\begin{tabular}{|c|c|c|c|}
%\hline
%NN$_{Higgs-signal}$ $\downarrow$ & \multicolumn{3}{|c|}{S$/$$\sqrt{B}$} \\ \hline
%NN $>$ 0.4 & 0.3 & 4.8 & 6.3 \\ \hline
%NN $>$ 0.2 & 0.4 & $\textbf{8.7}$ & 6.0 \\ \hline
%No cut & 0.1 & 0.7 & 1.0 \\ \hline 
%NN$_{SM-Higgs}$ $\rightarrow$ & No cut & NN $>$ 0.3 & NN $>$ 0.6 \\ \hline
%\end{tabular}
%\caption{Optimization of neural net cuts using S/$\sqrt{B}$ values in the hadronic channel.}
%\label{tab:4Nun1n2}
%\end{table}

\section{Branching ratio calculation and results}

The branching ratio of the Higgs boson decay to charm quarks was calculated using events that passed the final neural network 
selection. The calculation was done by normalising the signal cross section to the inclusive Higgs cross section, $\sigma_{ZH}$ = 209$\pm$9.8 fb, 
as determined in an independent recoil mass analysis performed for the SiD Letter of Intent~\cite{sidloi}. The branching ratio is then given by
\begin{equation}
BR(H \rightarrow f\bar{f}) = \frac{\sigma_{Hff}}{\sigma_{ZH}}
\end{equation}
where $f$ represents the daughter decay products from the Higgs.
The relative accuracy of the the branching ratio takes into account both the relative signal cross section uncertainty and
the relative Higgs-strahlung uncertainty given as
\begin{equation}
\frac{\Delta BR}{BR} = \sqrt{\left(\frac{\Delta\sigma_{Hff}}{\sigma_{Hff}}\right)^2 + \left(\frac{\Delta\sigma_{ZH}}{\sigma_{ZH}}\right)^2}
\end{equation}
with the relative signal cross section uncertainty calculated by
\begin{equation}
\frac{\Delta\sigma_{Hff}}{\sigma_{Hff}} = \frac{\sqrt{signal + background}}{signal}
\end{equation}
and the cross-section is calculated as follows
\begin{equation}
\sigma_{Hff} = \frac{N}{\varepsilon_{Hff} L}
\end{equation}
where N is the number of signal events after all selections, $\varepsilon$ is the efficiency of signal selection and L is the total integrated luminosity.

The uncertainty in the efficiency is considered negligible, relying on simulations to determine it with sufficient precision. The 
systematic effects or contributions of the luminosity uncertainty were not considered in this analysis.

The weighted average of the signal cross section and its uncertainty are calculated using cross section and relative uncertainty values obtained from the neutrino and hadronic channels. The weighted average cross section is given by
\begin{equation}
\sigma_{average} = \frac{x(\delta y)^2 + y(\delta x)^2}{(\delta x)^2 + (\delta y)^2},
\end{equation}
where $x$ and $y$ are the cross sections in the neutrino and hadronics channels respectively, and $\delta x$ and $\delta y$ are the cross section 
uncertainties in the neutrino and hadronic channels respectively assuming that the two channels are statistically independent. The uncertainty of the average cross section is then calculated as
\begin{equation}
\delta z = \frac{\delta x*\delta y}{\sqrt{(\delta x)^2 + (\delta y)^2}},
\end{equation}
where $\delta z$ is the uncertainty of the weighted average cross section.

A summary of the results obtained for both the neutrino and hadronic channels is given in Tables~\ref{tab:zhbr}
The relative accuracy of the branching ratio in both channels is dominated by the precision of determination of the signal cross section since the uncertainty of the total Higgs-strahlung cross section is small (4.7\%). The larger value of the BR uncertainty in the neutrino channel is largely due to the lower signal statistic since BR(Z $\rightarrow$ $\nu$$\bar{\nu}$) is 20.0\% ,and BR(Z $\rightarrow$ q$\bar{q}$) is 69.91\%.

\begin{table}[h]
\centering
\begin{tabular}{|l|l|l|l|l|}
\hline & Neutrino & Hadronic & Combined  \\
\hline
Signal events & 178 & 407 & \\ \hline
SM background events & 140 & 673 & \\ \hline
Higgs background events & 109 & 213 & \\ \hline
Signal efficiency \% & 27.943$\pm$0.018 & 22.156$\pm$0.010 & \\ \hline
Signal $\sigma_{Hcc}$ & 6.8$\pm$0.8 fb & 6.9$\pm$0.6 fb & 6.86$\pm$0.48 fb \\ \hline
Relative uncertainty on $\sigma_{Hcc}$ & 11.6\% & 8.8\% & 7.0\% \\ \hline
\end{tabular}
\caption{Measurement results of H$\rightarrow c\bar{c}$ branching ratio.}
\label{tab:zhbr}
\end{table}
Using Equation 5 the uncertainty on the branching ratio is found to be 8.4\%.

\section{Conclusion}

The uncertainty of the measurement of the decay branching ratio to charm quarks for a neutral SM Higgs boson of mass 120 GeV has been studied 
at a centre-of-mass of energy of $\sqrt{s}$ = 250 GeV and a total integrated luminosity of 250$\displaystyle\int$fb$^{-1}$. The analysis is based on full detector simulation and realistic event reconstruction. The relative uncertainties obtained are comparable to values obtained in some of the previous studies. A good performance of flavour tagging and the use of neural networks in event selection are critical in obtaining these results. In this study we find the uncertainty on the cross section to be 11.6\% in the neutrino channel and 8.8\% in the hadronic channel giving a combined uncertainty on the branching ratio of 8.4\%.  

\bibliographystyle{unsrt}

\end{document}